\documentclass[fleqn]{2023SCGE}
\usepackage{tabularx}
\usepackage{multirow}
\usepackage{graphicx} 
\usepackage{float} 
\usepackage{lineno}

\usepackage{cite}

\DeclareUnicodeCharacter{202C}{}
\begin{document}
\ensubject{subject}
\ArticleType{Article}
\SpecialTopic{SPECIAL TOPIC: }
\Year{2024}
\Month{January}
\Vol{66}
\No{1}
\DOI{??}
\ArtNo{000000}
\ReceiveDate{January 11, 1111}
\AcceptDate{April 1, 1111}
\OnlineDate{January 1, 1111}

\title{Detector performance of the Gamma-ray Transient Monitor onboard DRO-A Satellite}{Detector performance of the Gamma-ray Transient Monitor onboard DRO-A Satellite}

\author[1,2]{Pei-Yi Feng}{{fengpeiyi@ihep.ac.cn}}
\author[1]{Zheng-Hua An}{anzh@ihep.ac.cn}
\author[1]{Da-Li Zhang}{zhangdl@ihep.ac.cn}
\author[1,2]{Chen-Wei Wang}{}
\author[1,2]{Chao Zheng}{}
\author[1]{Sheng Yang}{}
\author[1]{\\Shao-Lin Xiong}{}
\author[1,2]{Jia-Cong Liu}{}
\author[1]{Xin-Qiao Li}{}
\author[1]{Ke Gong}{}
\author[1]{Xiao-Jing Liu}{}
\author[1]{Min Gao}{}
\author[1]{\\Xiang-Yang Wen}{}
\author[1]{Ya-Qing liu}{}
\author[1]{Xiao-Yun Zhao}{}
\author[1]{Fan Zhang}{}
\author[3]{Xi-Lei Sun}{}
\author[1]{Hong Lu}{}

\AuthorMark{Feng P Y}

\AuthorCitation{Feng P Y , An Z H, Zhang Z L, et al}

\address[1]{Key Laboratory of Particle Astrophysics, Institute of High Energy Physics, Chinese Academy of Sciences, Beijing 100049, China}
\address[2]{University of Chinese Academy of Sciences, Chinese Academy of Sciences, Beijing 100049, China}
\address[3]{State Key Laboratory of Particle Detection and Electronics, Institute of High Energy Physics, Chinese Academy of Sciences, Beijing 100049, China}

\abstract{

The Gamma-ray Transient Monitor (GTM) is an all-sky monitor onboard the Distant Retrograde Orbit-A (DRO-A) satellite with the scientific objective of detecting gamma-ray transients ranging from 20 keV to 1 MeV. The GTM was equipped with five Gamma-ray Transient Probe (GTP) detector modules utilizing a NaI(Tl) scintillator coupled with a SiPM array. To reduce the SiPM noise, GTP uses a dedicated dual-channel coincident readout design. In this work, we first studied the impact of different coincidence times on the detection efficiency and ultimately selected a 0.5 $\mu$s time coincidence window for offline data processing. To test the performance of the GTPs and validate the Monte-Carlo-simulated energy response, we conducted comprehensive ground calibration tests using the Hard X-ray Calibration Facility (HXCF) and radioactive sources, including the energy response, detection efficiency, spatial response, bias-voltage response, and temperature dependence. We extensively present the ground calibration results and validate the design and mass model of the GTP detector, thus providing the foundation for in-flight observations and scientific data analysis.

}


\maketitle

\begin{multicols}{2}
\section{Introduction}

On August 17, 2017, a large collaboration involving LIGO, Virgo, and more than 70 observatories reported the historical detection of a binary neutron star merger gravitational wave event (GW170817) and its corresponding gamma-ray burst (GRB170817A). Subsequently, their electromagnetic counterparts were identified across optical, soft X-ray,\Authorfootnote

\noindent and radio wavebands \cite{ref1, ref2, ref3, ref4}. Recent years have witnessed substantial advancements in gravitational waves (GWs), fast radio bursts (FRBs), high energy neutrinos (HENs), and cosmic rays (CRs), signifying the ‘multi-messenger multi-wavelength’ astronomy era \cite{ref5, ref6, ref7, ref8}. Since May 2023, ground-based gravitational wave detectors (LIGO, Virgo, and KAGRA) have initiated a new phase of scientific observation called O4. The high-energy electromagnetic counterparts associated with GWs have substantial discovery opportunities \cite{ref9, ref10, ref11, ref12}.

GECAM is a dedicated all-sky monitor constellation with the primary objective of detecting and localizing transient gamma-ray sources, particularly those associated with GWs and FRBs \cite{ref13, ref14, ref15, ref16, ref17, ref18}. Currently, three GECAM instruments, GECAM-A, GECAM-B, and GECAM-C (also called HEBS), have been successfully launched into low Earth orbit, leading to a series of discoveries \cite{ref19, ref20, ref21, ref22, ref41}. For example, GECAM-C accurately measured the brightest gamma-ray burst (GRB221009A) without any data saturation or other instrument-related artifacts \cite{ref23, ref24}.

As a new member of the high-energy astronomical transient monitoring network, already consisting of the GECAM series and Insight-HXMT, the GTM was designed and scheduled to be launched onboard the DRO mission in early 2024. The operational orbit of the DRO-A satellite ranges from approximately 310,000–450,000 km from the Earth and 50,000–100,000 km from the Moon. In contrast to detectors in low-Earth orbits, GTM in deep-space orbits offers notable advantages, including an instantaneous all-sky field of view (unobstructed by nearby celestial bodies) and a more stable space environment (without SAA passage) \cite{ref43}.

Considering the GTM largely inherits the hardware, software, and scientific operations of the GECAM mission, it is referred to as GECAM-D in the GECAM family. Similar to other GECAM instruments, the GTM has a real-time GRB trigger system based on K-band telemetry and the BeiDou short message service. The latter was first proposed and implemented during the GECAM missions (GECAM-B and GECAM-C) \cite{ref25}. Real-time GRB trigger alerts can inform other ground- and space-based telescopes to perform follow-up observations, facilitating multiwavelength joint observations of transients.
With GTM in deep space and other instruments in low Earth orbit, a high-energy astronomical transient monitoring network can be strengthened to monitor various high-energy transients, including GRBs, SGRs, and the high-energy electromagnetic counterparts of GWs and FRBs \cite{ref20, ref26, ref27}.

\begin{figure}[H]
\centering
\includegraphics[width=\columnwidth]
{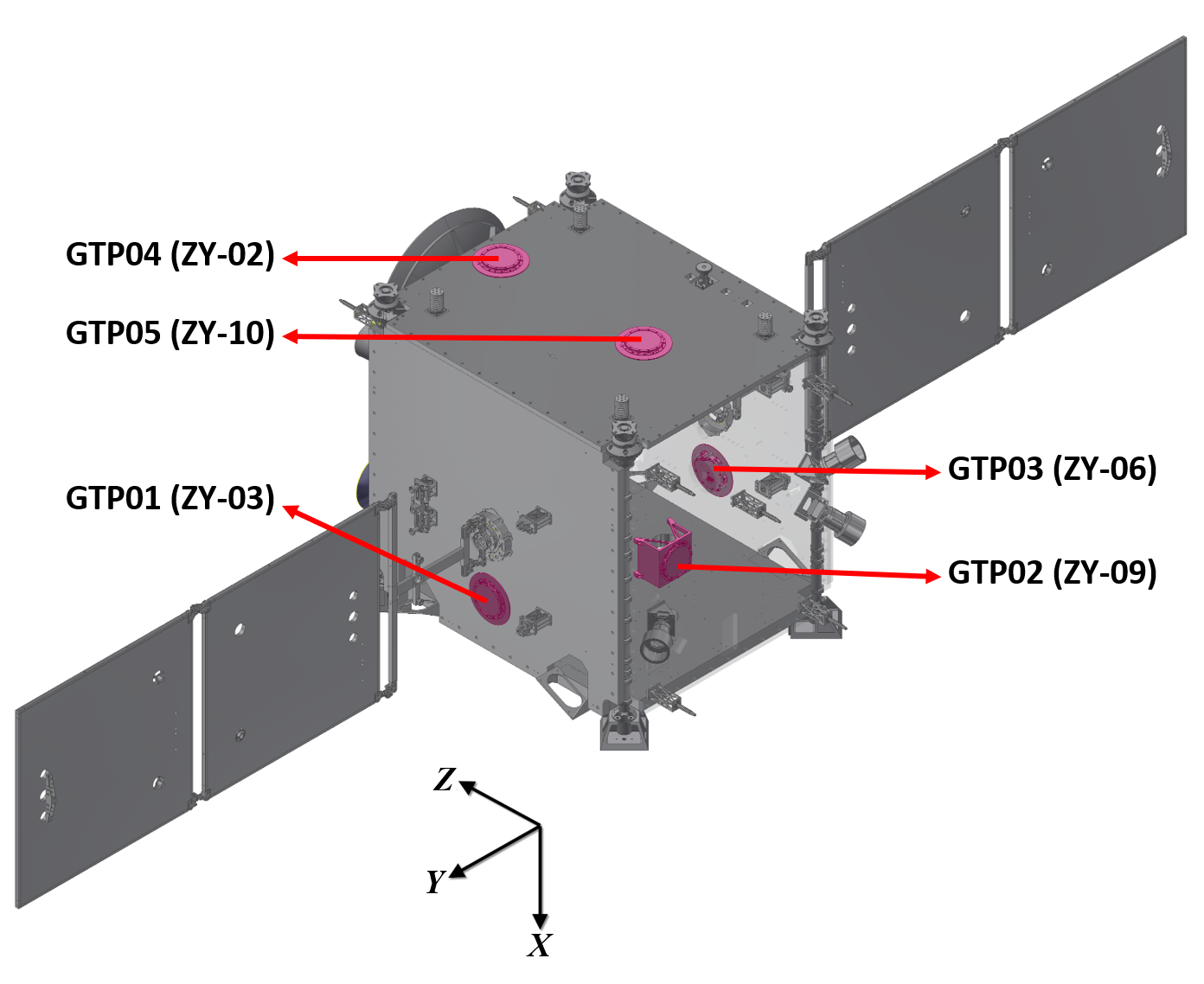}
\caption{Overview of DRO-A satellite. GTM consists of five Gamma-ray Transient Probes (GTPs) positioned on the four sides of the spacecraft. Four standard GTPs are individually mounted on the ±Y side (one GTP for each side) and the –X side (two GTPs), while one dedicated GTP for the –Z side. The standard GTP comprises detector components and radiation cooling plates, while the dedicated GTP is composed of detector components and brackets (with the brackets also doubling as radiation cooling plates). The detector label with GTP is for science usage, whereas that with ZY is for the crystal label of detector.}
\label{fig:figure 1}
\end{figure}

The GTM is equipped with five GTPs, primarily designed for detecting gamma rays in the energy range of 20 keV–1 MeV and for the rough localization of gamma-ray transient sources. Each GTP had an approximately 2$\pi$ field of view, with the five detectors placed in various orientations on the surface of the DRO-A satellite with partial overlap in their fields of view, ultimately achieving nearly all-sky coverage (Fig.~\ref{fig:figure 1}). To accurately measure the spectral information of gamma-ray transient sources, conducting thorough ground calibrations of GTPs, including the time coincidence, energy response, detection efficiency, spatial response, bias-voltage response, and temperature dependence, is important. Ground calibration of the GTM was also employed to validate the GTP energy-response matrix derived through Monte Carlo simulations based on satellite and detector mass models. Ground calibration results can also serve as inputs for in-flight calibration and integrating both key components of the calibration database.

\section{Instrument Design and Ground Calibration}

This section covers the design of the GTP, ground calibration setup for hard X-rays, and types of radioactive sources required for calibration.

\subsection{Design of Gamma-ray Transient Probe}

\begin{figure}[H]
\centering
	\includegraphics[width=\columnwidth]{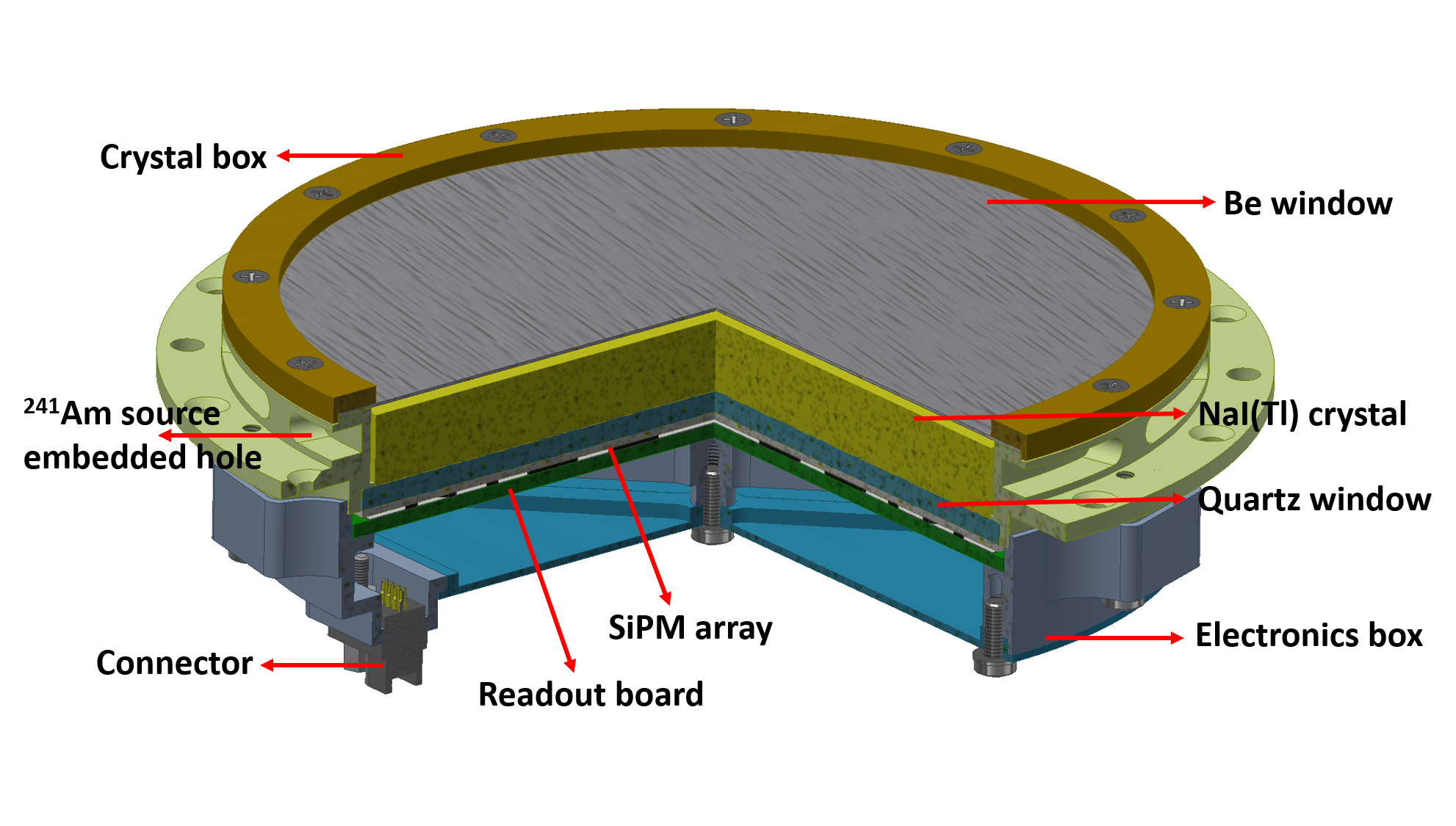}
    \caption{Structural diagram of Gamma-ray Transient Probe (GTP) module onboard GTM.}
    \label{figure 2}
\end{figure}

\begin{figure}[H]
\centering
	\includegraphics[width=\columnwidth]{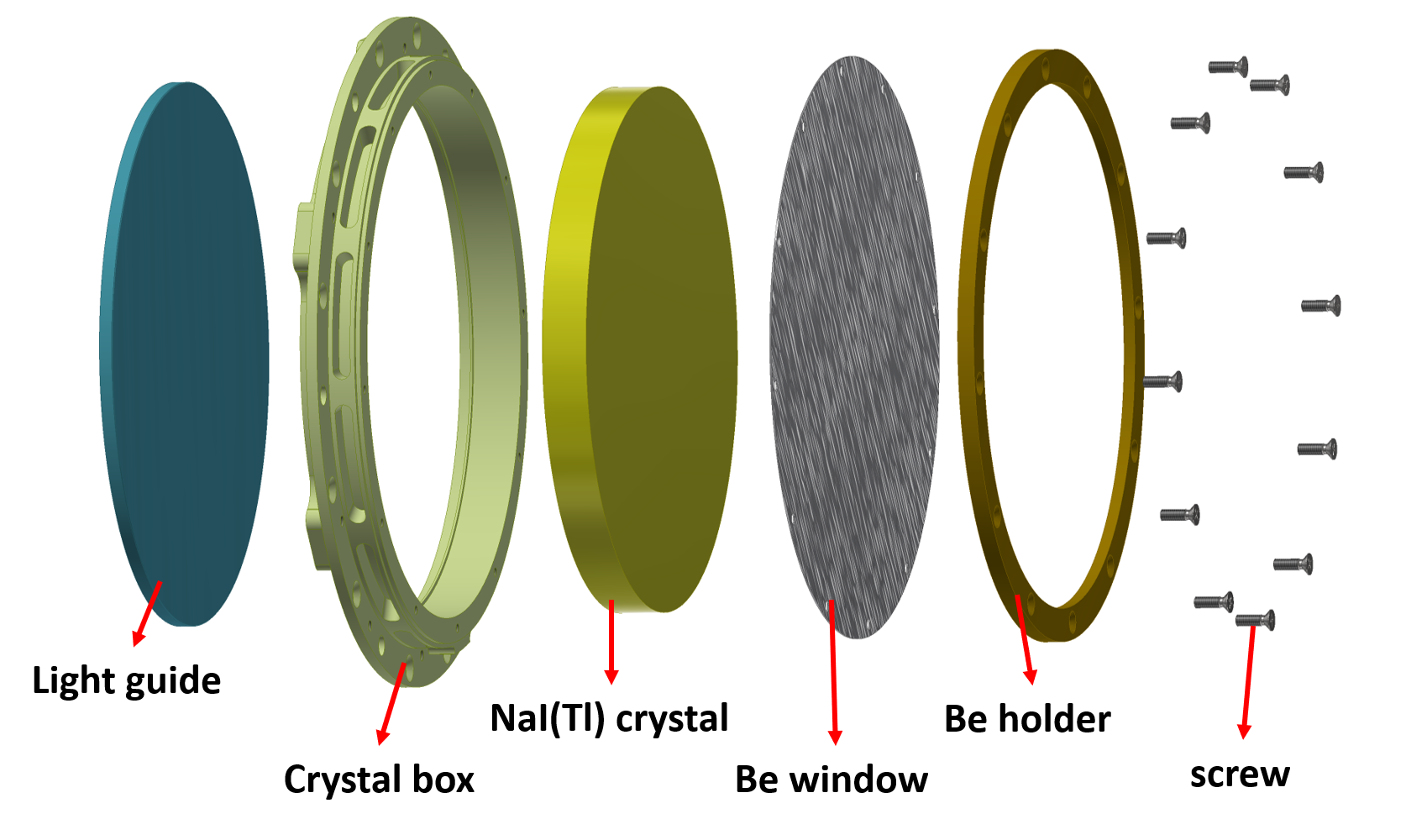}
    \caption{Schematic diagram of NaI(Tl) crystal packaging.}
    \label{figure 3}
\end{figure}

Similar to the design of the gamma-ray detector (GRD) in GECAM-A/B and GECAM-C \cite{ref23, ref28, ref42}, the structure of the GTM GTP is depicted in Fig.~\ref{figure 2}. The GTP employs a NaI(Tl) crystal with a diameter of 115 mm and a thickness of 10 mm, manufactured by Beijing Glass Research Institute Co., Ltd., as its sensitive detection material. 
The NaI(Tl) crystal requires fully sealed packaging, as shown in Fig.~\ref{figure 3}. The incident window was made of a 400-$\mu$m-thick Be sheet with a transmission rate greater than 80\% for 10 keV gamma rays. The light output window was equipped with a 3 mm-thick quartz glass and coupled to a 100-chip SiPM array (MICROFJ-60035-TSV-TR) through a 1 mm-thick optical silicone rubber for readout. The NaI(Tl) crystal was coated with two layers of 0.1 mm-thick and two layers of 0.2 mm-thick PTFE (Teflon). Teflon was placed around the NaI(Tl) crystal and on its top primarily to reflect the scintillation photons, thereby increasing the collection efficiency of the scintillation photons.
GTP simultaneously provides energy and time information for the gamma rays used for physical analysis. We designed a hole on the side of the GTP to install a $^{241}$Am radioactive source to facilitate in-orbit calibration. 
These $^{241}$Am sources have a diameter of 2.5 mm and a thickness of 3 mm, with an activity range of 500 to 800 Bq. The primary technical characteristics of the GTP are listed in Table~\ref{tab:characteristics of the GTP}.

\begin{table}[H]
	\centering
	\caption{Main characteristics of the flight GTP detectors of GTM.}
	\label{tab:characteristics of the GTP}
	\begin{tabular*}{8.5cm} {@{\extracolsep{\fill} } >{\centering\arraybackslash}m{4cm}  >{\centering\arraybackslash}m{4cm} } 
		\hline
		GTP Parameters & Value\\
		\hline
		Type & NaI(Tl)+SiPM\\
		Detector Area & 103.87 cm$^2$\\
            Energy range & 20–1000 keV\\
            Energy resolution & $\leq$ 25\%@59.5 keV\\
            Detection efficiency & $\geq$ 60\%@20 keV\\
            Deadtime & 4 $\mu$s (normal)\\
		\hline
	\end{tabular*}
\end{table}

\begin{figure}[H]
\centering
	\includegraphics[width=0.8\columnwidth]{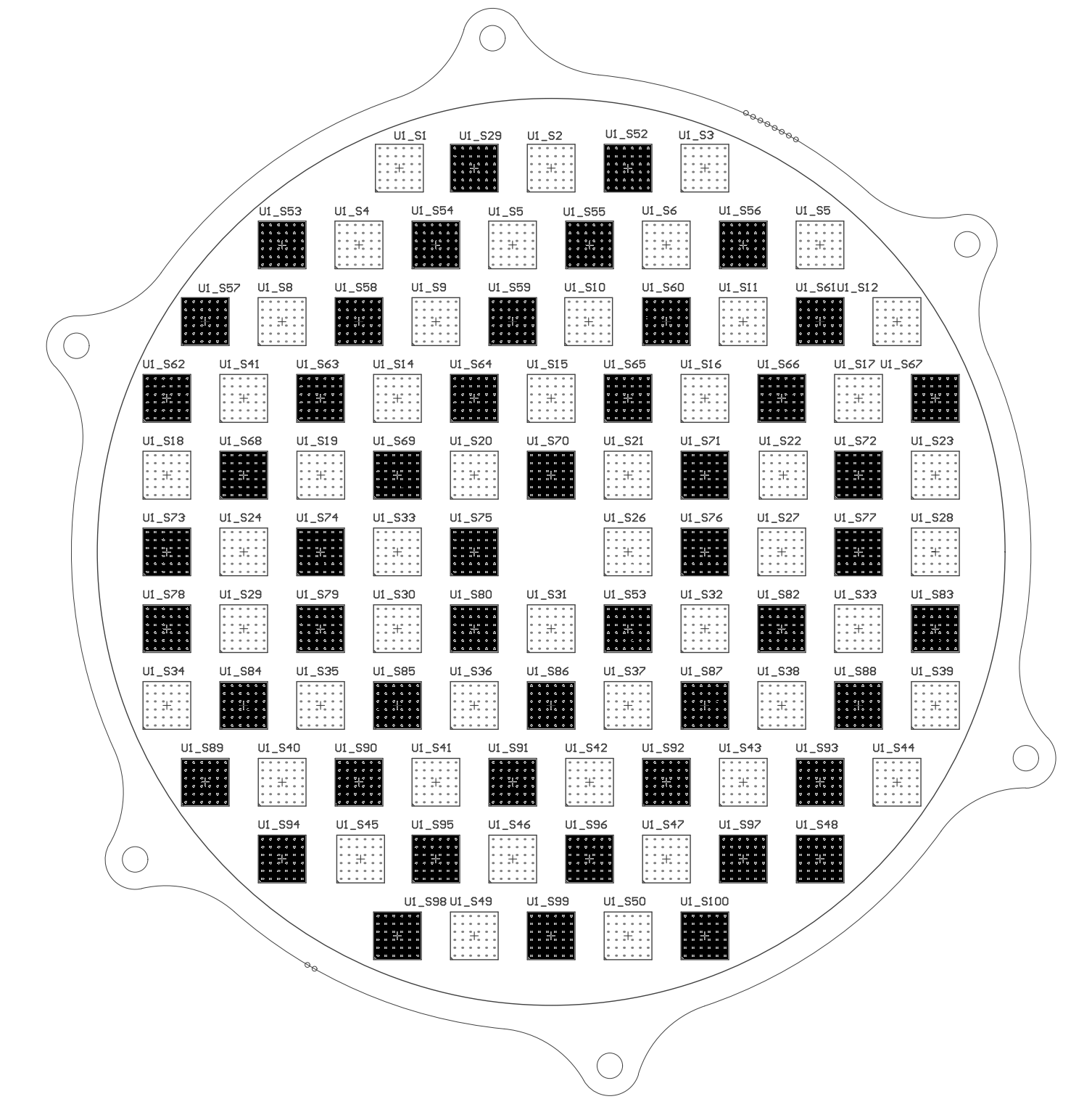}
    \caption{Schematic diagram of SiPM array grouping. White blocks represent group 1, black blocks represent group 2.}
    \label{figure 4}
\end{figure}

\begin{figure}[H]
\centering
	\includegraphics[width=\columnwidth]{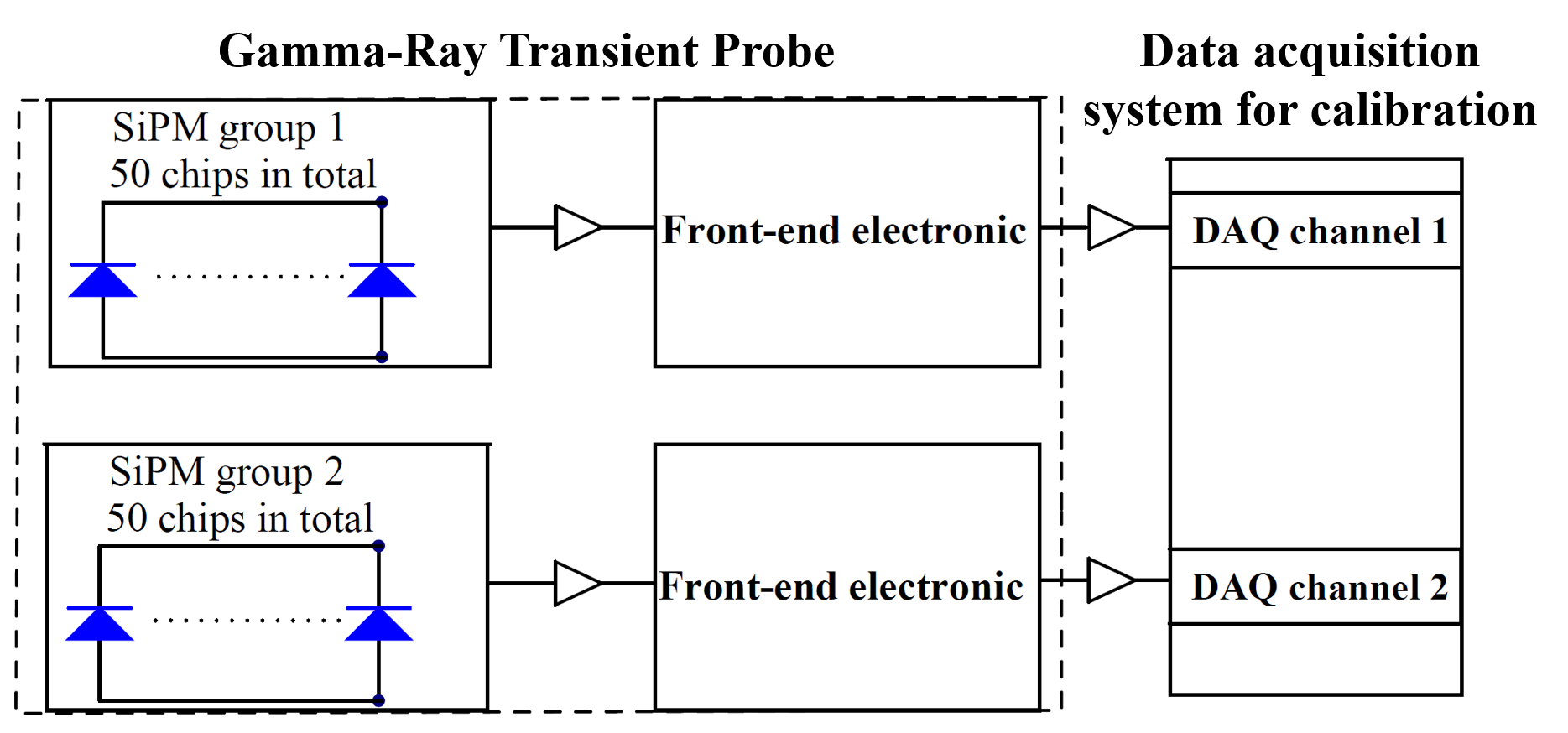}
    \caption{Block diagram of read-out electronics and data acquisition system.}
    \label{figure 5}
\end{figure}

Figure~\ref{figure 4} shows the uniform division of the SiPM array within the GTP into two groups. Each group underwent front-end electronic readout (Fig.~\ref{figure 5}) and the resulting signal was acquired using a data acquisition system for calibration purposes. Each SiPM group's output is connected in parallel, and the two groups have an independent readout. Offline data processing involves DAQ channels 1 and 2 with each event data, including signal amplitude and timestamp information. The timestamps facilitated time-coincidence data analysis to mitigate SiPM dark noise, utilizing a 0.5 $\mu$s coincidence window in this ground calibration test. The final data acquisition system for the GTM payload was designed using a dedicated real-time signal coincidence method, which will be described in detail in future papers.

The GTM comprises 10 GTPs, which include five flight GTPs (designated as ZY-02, ZY-03, ZY-06, ZY-09, and ZY-10), two backup GTPs (designated as ZY-01 and ZY-08), two test GTPs (designated as ZY-04 and ZY-07), and one assessment GTP (designated as ZY-05). We conducted a comprehensive ground calibration for these GTPs; the calibration items and their corresponding GTP designations are listed in Table~\ref{tab:calibrationproject}. Ground calibrations for the energy responses of all the GTPs were performed at the National Institute of Metrology (NIM). HXCF was employed to calibrate the low-energy range (9–160 keV), and a series of radioactive sources were used for the high-energy range.

\begin{table}[H]
	\centering
	\caption{Ground calibration items and their corresponding Gamma-ray Transient Probe (GTP) designations.}
	\label{tab:calibrationproject}
	\begin{tabular*}{8.5cm} {@{\extracolsep{\fill} } >{\centering\arraybackslash}m{3.6cm}  >{\centering\arraybackslash}m{4.5cm} } 
		\hline
		Calibration project & GTP identification number\\
		\hline
		Energy response & ZY-01, ZY-02, ZY-03, ZY-06, ZY-08, ZY-09, ZY-10\\
		Detection efficiency & ZY-01, ZY-02, ZY-03, ZY-06, ZY-08, ZY-09, ZY-10\\
		Position response & ZY-01, ZY-02, ZY-03, ZY-06, ZY-08, ZY-09, ZY-10\\
            Bias-voltage response & ZY-06, ZY-08\\
            Temperature dependence & ZY-05 \\
		\hline
	\end{tabular*}
\end{table}

\subsection{Ground Calibration with the Hard X-ray Calibration Facility}

\begin{figure}[H]
\centering
	\includegraphics[width=\columnwidth]{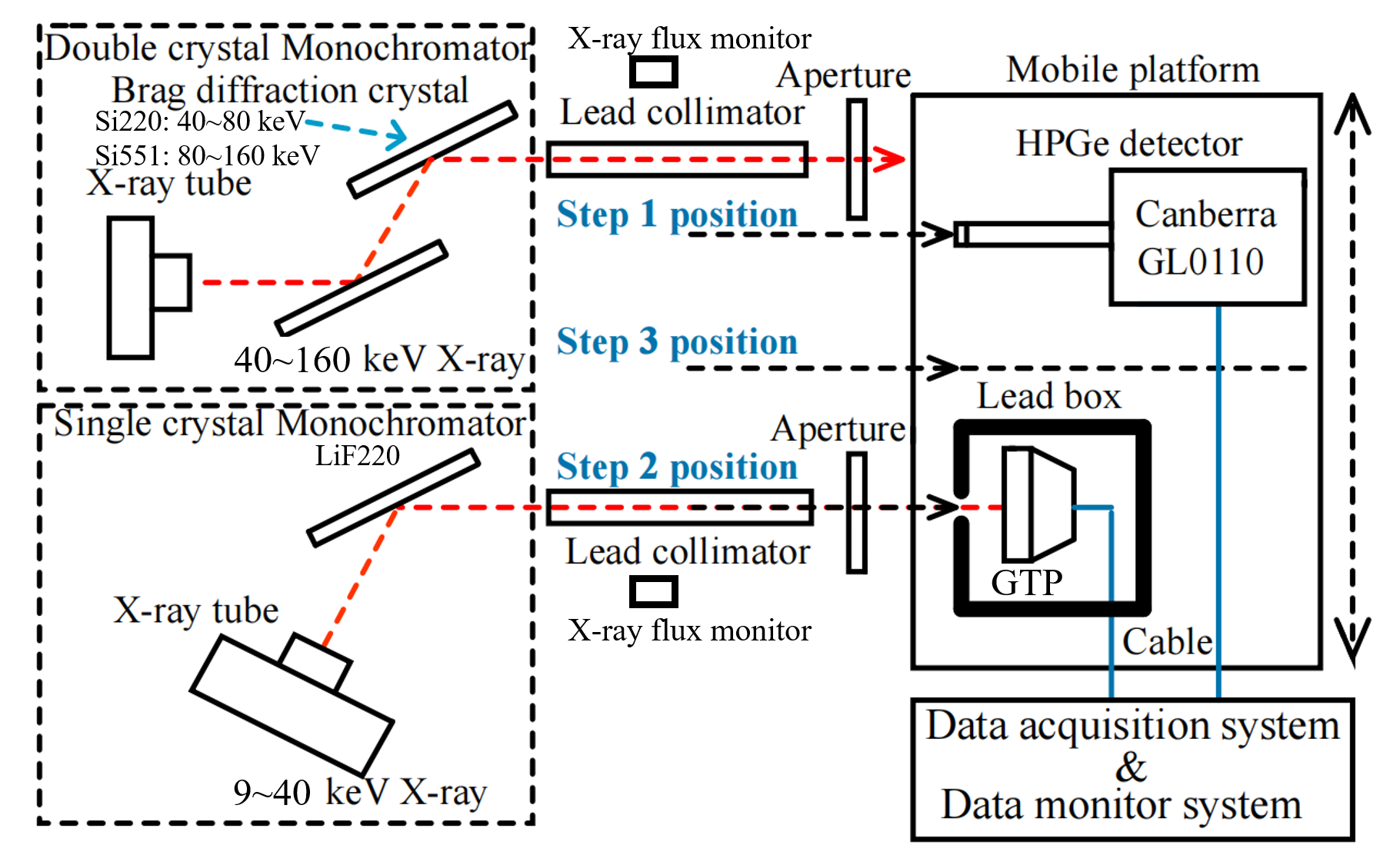}
    \caption{Hard X-ray Calibration Facility (HXCF). A lanthanum bromide (LaBr$_3$) crystal was employed as the X-ray flux monitor.}
    \label{figure 6}
\end{figure}

The energy response of the GTP to X-rays in the 9–160 keV range was calibrated using a Hard X-ray Calibration Facility (HXCF) at the NIM in Changping, Beijing, China \cite{ref29, ref30, ref31}. Originally established for the high-energy Hard X-ray Modulation Telescope (HXMT), HXCF has also played a substantial role in ground calibrations for gamma-ray detectors of GECAM-A/B, GECAM-C, and Space Variable Object Monitor (SVOM) satellites \cite{ref32, ref33, ref34, ref35}. HXCF comprises an X-ray generator, a single-crystal monochromator (9–40 keV), a double-crystal monochromator (40–160 keV), a collimation system to suppress stray light and limit X-ray beam size, a displacement platform for positioning detectors at X-ray beam locations, a lead shielding system, an X-ray flux monitor, and a well-calibrated High-Purity Germanium (HPGe) detector (GL0110P manufactured by Canberra) for precise measurement of X-ray energy and flux (Fig.~\ref{figure 6}) \cite{ref23}.

\begin{figure}[H]
\centering
	\includegraphics[width=\columnwidth]{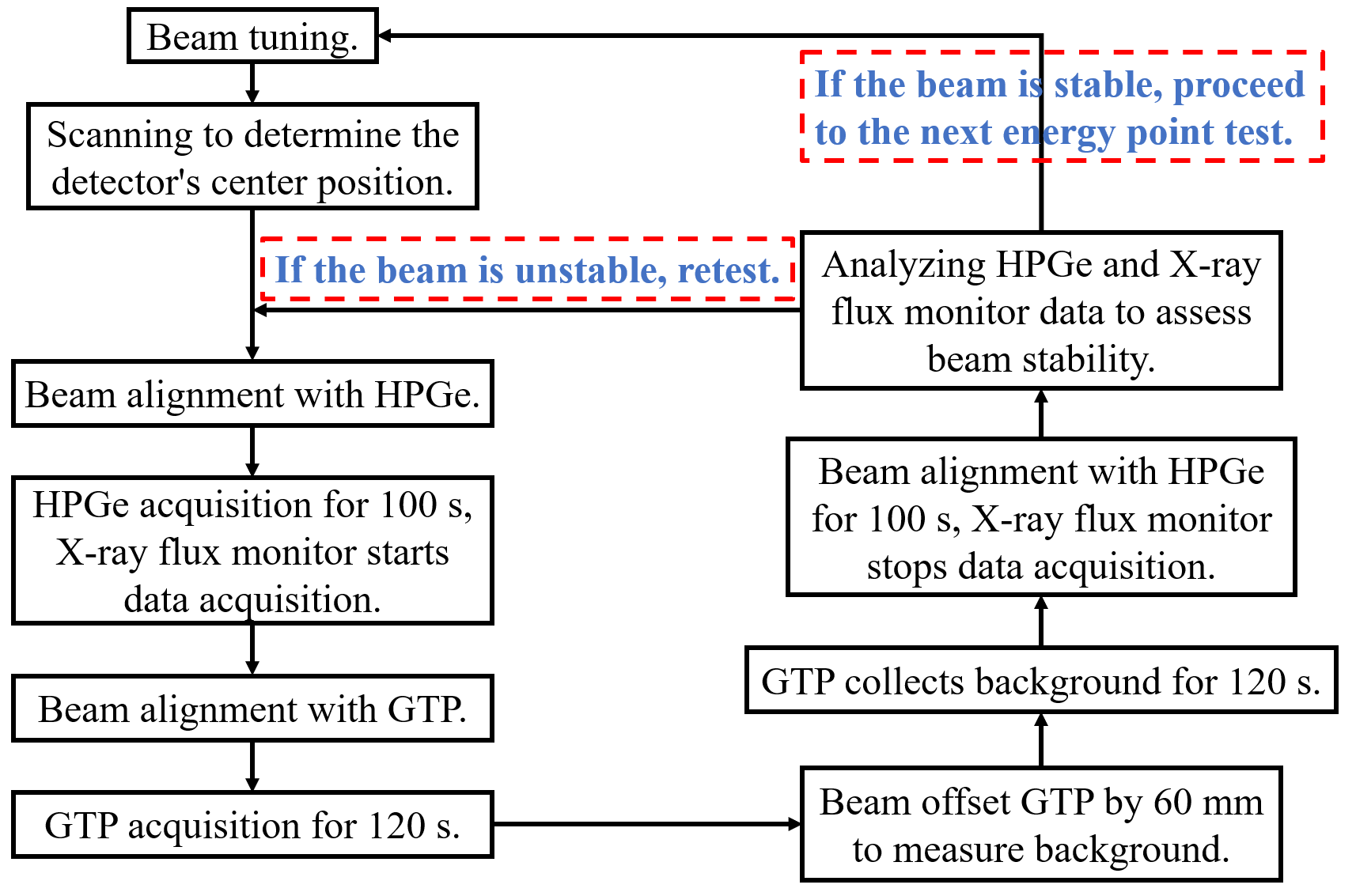}
    \caption{Flowchart of the GTP ground calibration procedures using the Hard X-ray Calibration Facility (HXCF).}
    \label{figure 7}
\end{figure}

To shield against the impact of scattered and leaked X-rays in the environment, the GTP was positioned inside a 10 mm-thick lead enclosure placed alongside the HPGe detector on a displacement platform. The shielding box had a small aperture with a diameter of 10 mm in the X-ray beam direction. This aperture size is similar to that of the HPGe detector, and it exceeds the 3 mm diameter of the X-ray beam spot, ensuring the precision of the GTP efficiency testing. The GTP ground calibration procedure is illustrated in Fig.~\ref{figure 7}. Each GTP was calibrated with a minimum of 30 energy points using an X-ray beam. During the ground calibration process, the ambient temperature was meticulously maintained at 20±2 ℃, and the SiPM bias voltage was fixed at 26.5 V. 
There are differences between the in-orbit temperature and ground calibration. The specific effects of the temperature and bias voltage on the response are further explained in Sections \ref{Bias-voltage Response} and \ref{Temperature Dependence}.

\subsection{Ground Calibration with the Radioactive Sources}

The energy response calibration for the GTPs in the high-energy range was conducted using a series of radioactive sources, as detailed in Table~\ref{tab:radioactivesources}. Because of the unevenness in the NaI(Tl) crystal luminescence and light collection, the distance between the radioactive source and GTP affected the energy response, with the results becoming consistent beyond a distance of 5 cm. Prior to each GTP test, we gathered 300 s of background data, followed by sequential placement of the radioactive source at a distance of 10 cm from the center of the GTP for data acquisition. 
In spectrum testing, it is essential to ensure a sufficient level of statistics to reduce statistical errors (statistics $\textgreater$ 10000). Table~\ref{tab:radioactivesources} presents the activity of this batch of radioactive sources. To maintain consistent detection statistics, the testing time for each source is determined by its activity.
Throughout the ground calibration process using sources, the laboratory temperature was meticulously maintained at 20 $\pm$ 2 ℃. Furthermore, spatial response calibration and temperature experiments for the GTPs were performed using an $^{241}$Am source, and bias-voltage response calibration was performed using both $^{241}$Am and $^{137}$Cs sources.

\begin{table}[H]
	\centering
	\caption{Properties of radioactive sources used for the ground calibration of Gamma-ray Transient Probes (GTPs).}
	\label{tab:radioactivesources}
	\begin{tabular*}{8.5cm} {@{\extracolsep{\fill} } >{\centering\arraybackslash}m{1cm}c>{\centering\arraybackslash}m{1.5cm}>{\centering\arraybackslash}m{1.5cm}c } 
 
		\hline
		Source & Half-life & Activity (Bq) & Energy (keV) & Intensity\\
		\hline
		$^{241}$Am & 432.2 y & 9.45$\times$10$^3$ & 59.54 & 35.78\%\\
		$^{57}$Co & 271.74 d & 2.865$\times$10$^4$ & 122.06 & 85.51\%\\
		\multirow{2}{*}{$^{133}$Ba} & \multirow{2}{*}{10.52 y} & \multirow{2}{*}{1.589$\times$10$^5$} & 356.01 & 62.05\%\\
         & & & 81 & 34\%\\
            $^{137}$Cs & 30.17 y & 9.495$\times$10$^3$ & 661.6 & 84.99\%\\
            \multirow{2}{*}{$^{22}$Na} & \multirow{2}{*}{2.6 y} & \multirow{2}{*}{5.086$\times$10$^4$} & 511 & 179.79\%\\
             & & & 1270 & 99.94\%\\
            $^{210}$Pb & 22.2 y & 7.623$\times$10$^3$ & 10.8 & 25\%\\
            \multirow{2}{*}{$^{152}$Eu} & \multirow{2}{*}{13.54 y} & \multirow{2}{*}{6.36$\times$10$^4$} & 40.1 & 28.36\%\\
             & & & 121.78 & 28.41\\
            $^{109}$Cd & 461.4 d & 9.807$\times$10$^4$ & 22.16 & 55.2\%\\
		\hline
	\end{tabular*}
\end{table}

\section{Ground Calibration Results}

This section presents the ground calibration results, including the time coincidence, energy response, detection efficiency, spatial response, bias-voltage response, and temperature dependence.

\subsection{Time Coincidence}

\begin{figure}[H]
\centering
	\includegraphics[width=\columnwidth]{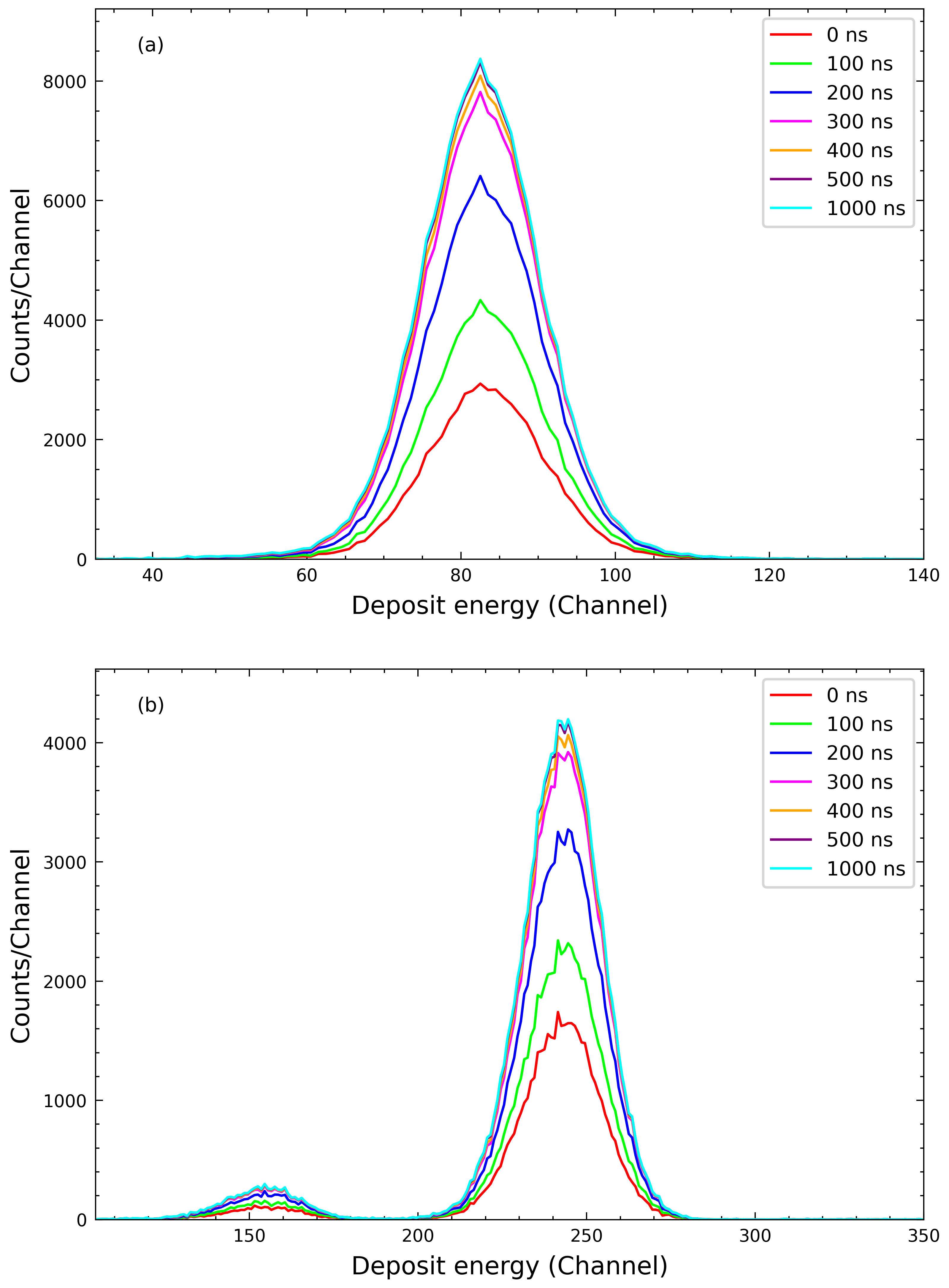}
    \caption{The impact of time coincidence on the counting rate in the offline processing of ZY-09 GTP's 25 keV (a) and 80 keV (b) X-ray data is reflected in the spectrum.}
    \label{figure 8}
\end{figure}

\begin{figure}[H]
\centering
	\includegraphics[width=\columnwidth]{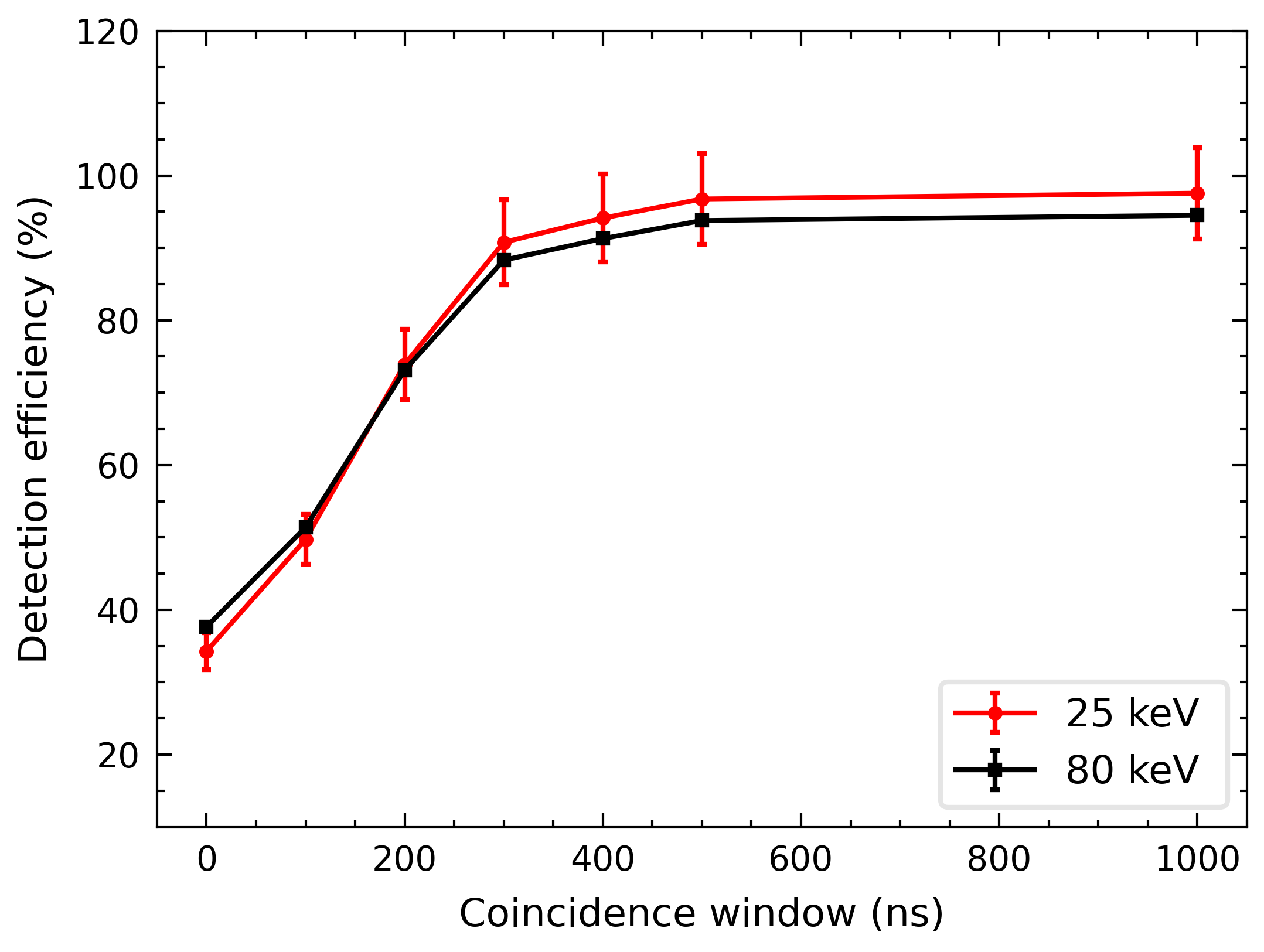}
    \caption{The impact of time coincidence on the detection efficiency for ZY-09 GTP's 25 keV and 80 keV X-ray data.}
    \label{figure 9}
\end{figure}

The SiPMs in GTPs adopt a grouped design, read signals from dual channels, and perform time coincidence, marking a significant innovation in GTM. To establish the optimal coincidence window, we investigated the ZY-09 GTP spectra (Fig.~\ref{figure 8}), and detection efficiency (Fig.~\ref{figure 9}) for 25 keV and 80 keV X-rays across different time-coincidence windows. 
Because the decay time of the NaI(Tl) crystal was 230 ns, along with the effects of electronic jitter and pre-amplification, we observed that when the coincidence time was less than 0.3 $\mu$s, events were lost, which significantly affected the detection efficiency. Conversely, when the window exceeded 0.5 $\mu$s, both the spectrum and detection efficiency remained relatively stable. However, excessively long coincidences impede noise elimination. For the ground calibration discussed in this paper, the time coincidence width was set at 0.5 $\mu$s. At low temperatures, the NaI(Tl) pulse widened slightly and the rise time slowed slightly. Although there was an impact, the adopted 0.5 $\mu$s was appropriate. Hence, we recommend setting the coincidence width between 0.5–1 $\mu$s.

\subsection{Energy Response}\label{chap:Energy Response}

\subsubsection{Energy Spectrum and Fitting}\label{chap:Energy Spectrum and Fitting}

\begin{figure}[H]
\centering
	\includegraphics[width=\hsize]{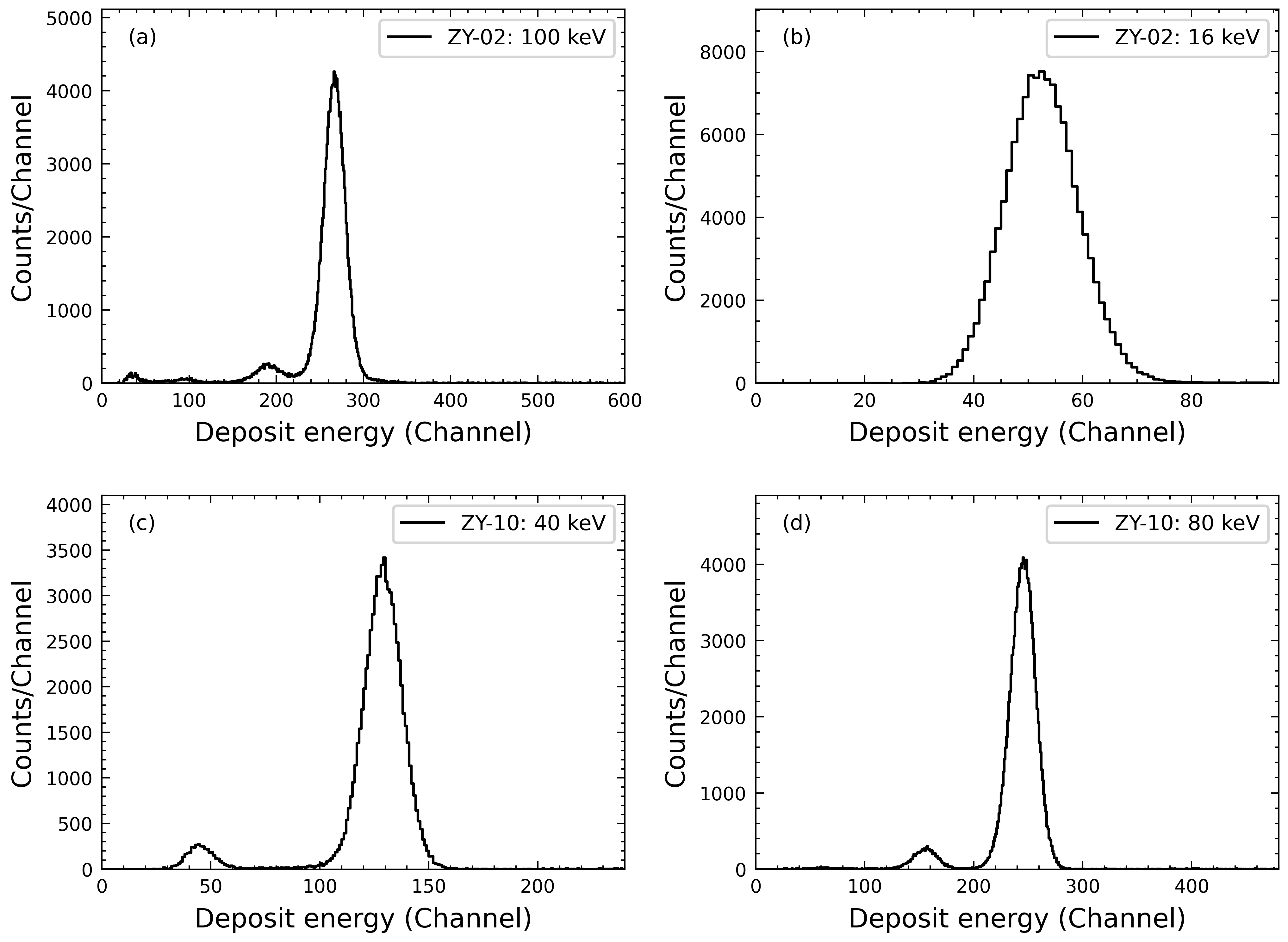}
    \caption{Net energy spectra of ZY-02 and ZY-10 GTPs were obtained using X-rays with energies of 100 keV (a), 16 keV (b), 40 keV (c), and 80 keV (d). These spectra underwent dead-time correction and background subtraction.}
    \label{figure 10}
\end{figure}

\begin{figure}[H]
\centering
	\includegraphics[width=\hsize]{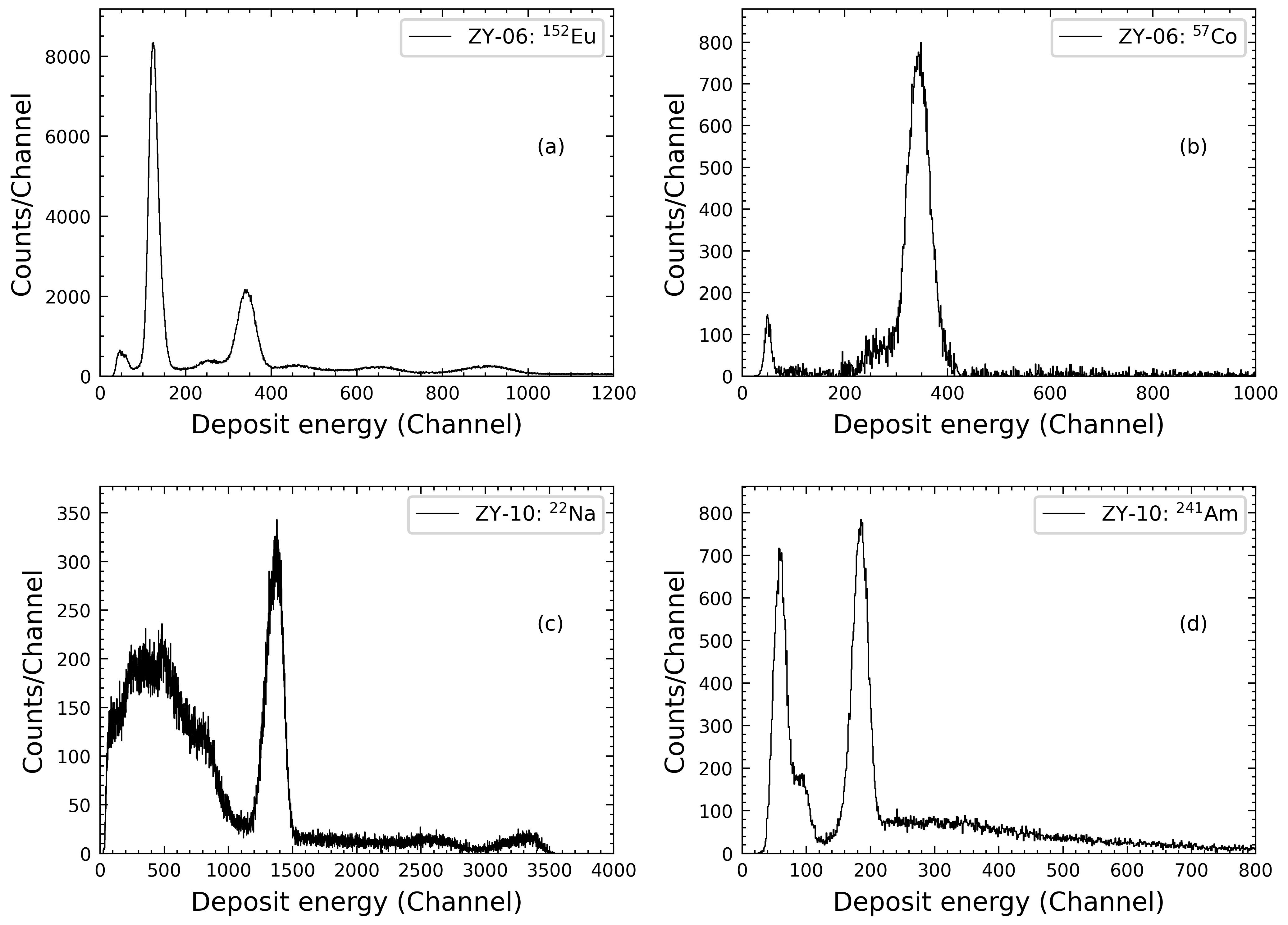}
    \caption{Net energy spectra of ZY-02 and ZY-10 GTPs were obtained using $^{152}$Eu (a), $^{57}$Co (b), $^{22}$Na (c), and $^{241}$Am (d) radioactive sources. These spectra were subjected to dead-time correction and background subtraction.}
    \label{figure 11}
\end{figure}

As the background varies with the testing environment, we acquire energy spectrum and background spectrum for the same duration. After applying dead time correction to both, we subtract the background spectrum from the energy spectrum to obtain the filtered spectrum. The filtered spectrum, corrected for dead time and background subtraction, is referred to as the net spectrum.
To obtain the net energy spectra of X-rays in the GTPs, we conducted dead-time correction and background subtraction for each spectrum using Equation~\ref{eq:background subtraction}. The partial net energy spectra of ZY-02 and ZY-10 are shown in Fig.~\ref{figure 10}. After processing the radioactive source spectra in the same manner, the partial net energy spectra of ZY-06 and ZY-10 were obtained, as shown in Fig.~\ref{figure 11}. 
The detection efficiency of GTP for the 511 keV gamma-ray full-energy peak of $^{22}$Na was only 11.71\%, resulting in relatively low statistics in the $^{22}$Na spectrum depicted in Fig.~\ref{figure 11}. However, the full-energy peak remained prominent, satisfying the requirement of statistical fluctuations of less than 1\%.

\begin{equation}\label{eq:background subtraction}
Pha_{net} = (\frac{Pha_X} {T_X - T_{dead,X}} - \frac{Pha_{bg}} {T_{bg} - T_{dead,bg}}) \cdot T_X.
\end{equation}

\begin{figure*}
\centering
	\includegraphics[width=\hsize]{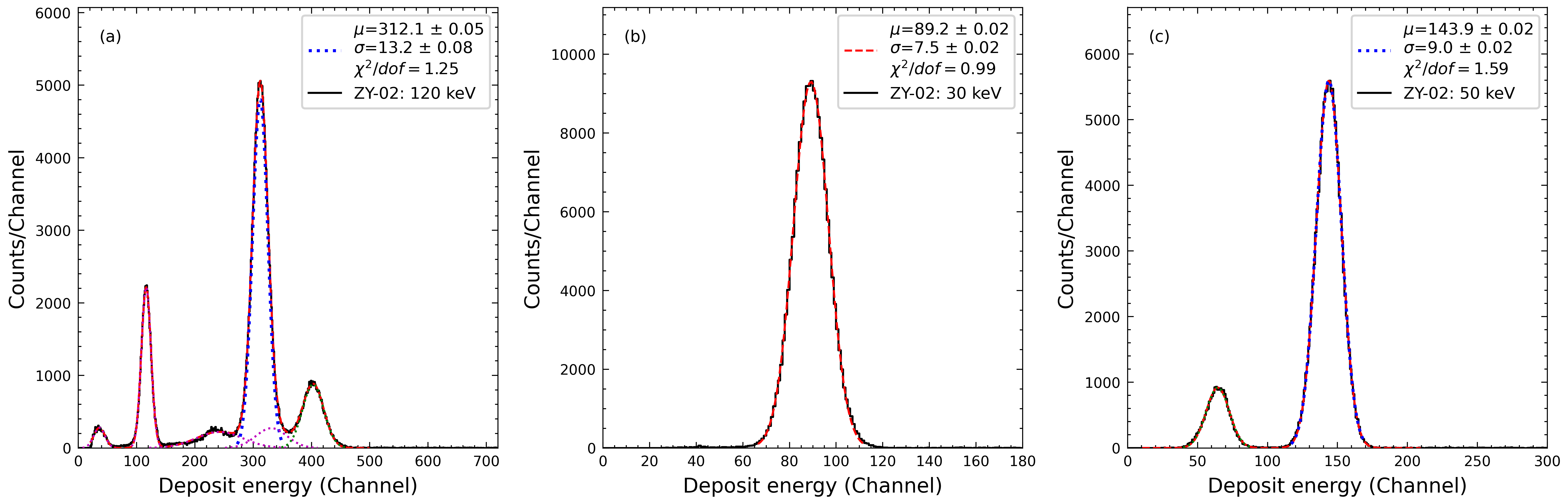}
    \caption{Multi-Gaussian fitting results of the net energy spectra for 120 keV (a), 30 keV (b), and 50 keV (c) X-rays by ZY-02 GTP.}
    \label{figure 12}
\end{figure*}

The X-ray full-energy peaks were accurately fitted and the corresponding fitting curves are shown in Figs.~\ref{figure 12}. These fitting results, comprising the centroid of the energy peak $\mu_i$, standard deviation $\sigma_i$, full width at half maximum ($FWHM = 2.355 \cdot \sigma_i$), and goodness of fit (represented by $\chi_i^2/dof$), provide crucial information for the energy response calibration of the GTPs. The radioactive source spectra are more complex than the X-ray tube spectra owing to the presence of Compton scattering and other decay types. Consequently, when fitting radioactive source spectra, it is necessary to incorporate additional functions, such as linear, quadratic, or exponential functions, to accommodate non-photon-peak contributions.

In Equation~\ref{eq:background subtraction}, $Pha_{net}$ represents the X-ray net energy spectrum, $Pha_X$ denotes the original X-ray energy spectrum, $Pha_{bg}$ signifies the background spectrum, $T_X$ and $T_{bg}$ correspond to the total testing time for the X-ray energy spectrum and background, while $T_{dead,X}$ and $T_{dead,bg}$ represent the dead times for the X-ray energy spectrum and background, respectively. 
The dead time of GTP is defined as the time from the signal crossing the threshold to the signal returning to the baseline. For normal events, the pulse width is approximately 2 $\mu$s. To ensure the signal returns to the baseline, the data acquisition system fixes the dead time for normal events at 4 $\mu$s. Events with pulse width greater than 4 $\mu$s are classified as overflow events. These events do not record amplitude but only record the length of the dead time. Corrections are made based on the recorded dead time.

When the X-ray energy decreased below the binding energy of the I's K-shell electrons (33.17 keV), a single full-energy peak appeared in the spectrum. However, for X-ray energies surpassing the binding energy of the I's K-shell electrons and where the remaining energy exceeded the threshold, an escape peak emerged on the left side of the full-energy peak. 
During ground calibration in the 80–160 keV energy range, Si551 crystals were employed as the HXCF monochromator. The X-rays incident on the two Si551 crystals resulted in Bragg diffraction. In Fig.~\ref{figure 6}, we have indicated that the crystal monochromator used for 80–160 keV are Si551 crystals, while for 40–80 keV, they are Si220 crystals, and for 9–40 keV, it is LiF200 crystal.

\subsubsection{Energy–Channel Conversion}

\begin{figure}[H]
\centering
	\includegraphics[width=\hsize]{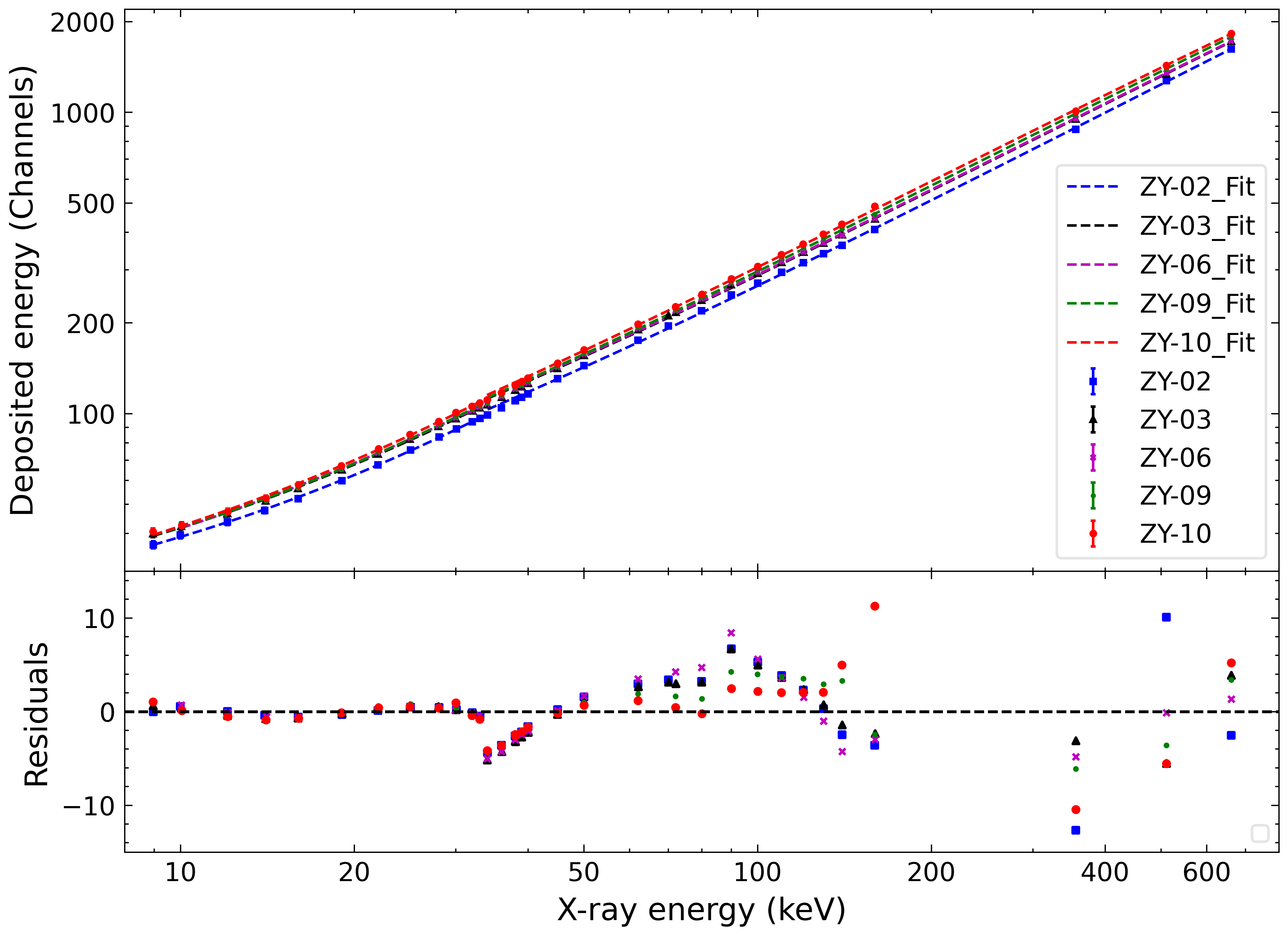}
    \caption{Energy–channel relationships of five flight GTPs were established by fitting these data points with the quadratic polynomial. The residuals were obtained by subtracting the experimental values from the fitted ones.}
    \label{figure 13}
\end{figure}

Based on the results of the Energy Spectrum and Fitting in Section~\ref{chap:Energy Spectrum and Fitting}, the Energy-Channel (E-C) relationships shown in Fig.~\ref{figure 13}. The binding energy of I's K-shell electrons (33.17 keV) served as the breakpoint. 
We used the binding energy of I's K-shell electrons (33.17 keV) serves as the breakpoint. We employed Equation~\ref{eq:E-C Fit} to perform segmented fitting on the data points ranging from 9 to 33.17 keV and from 33.17 keV to 662 keV \cite{ref35}. The fitting curves and residuals are shown in Fig.~\ref{figure 13}.

\begin{equation}\label{eq:E-C Fit}
Ch(E_\gamma) = b_0 + b_1x + b_2x^2.
\end{equation}

The presence of absorption edges led to relatively large residuals near 33.17 keV. Additionally, for the energy range 80–160 keV, the use of a double-crystal monochromator in HXCF results in multiple peaks, and fitting errors can increase the residuals. Furthermore, different testing methods, such as X-ray machine testing for low to medium energies (9–160 keV) and radioactive source testing for high energies ($\textgreater$160 keV), also affect the residuals in Fig.~\ref{figure 13}. The X-ray machine had a beam diameter of only 3 mm and was detected mainly in the central part of the GTP. During radioactive source testing, GTP detects nearly the entire surface, and the superposition of uniformity affects the energy response.

Considering the detection threshold and dynamic baseline along with the E-C relationship, the energy range can be determined. The energy detection range for all flight GTPs ranged from 9 to 1100 keV.
In fact, amplifiers or electronics exhibit non-linear behavior. The linear output range of the preamplifiers was $\pm$3 V. The nonlinear range was $\pm$(3–3.3~V). However, the ADC conversion range of read-out electronics is 0–2 V, which is the linear range of preamplifiers. In this manner, the ADC-to-keV relationship is also fully linear up to higher energies. If the signal amplitude is above the ADC range, this event will be marked as an ‘overflow’ event and will only record the deadtime of this event.

\subsubsection{Energy Resolution}

\begin{figure}[H]
\centering
	\includegraphics[width=\hsize]{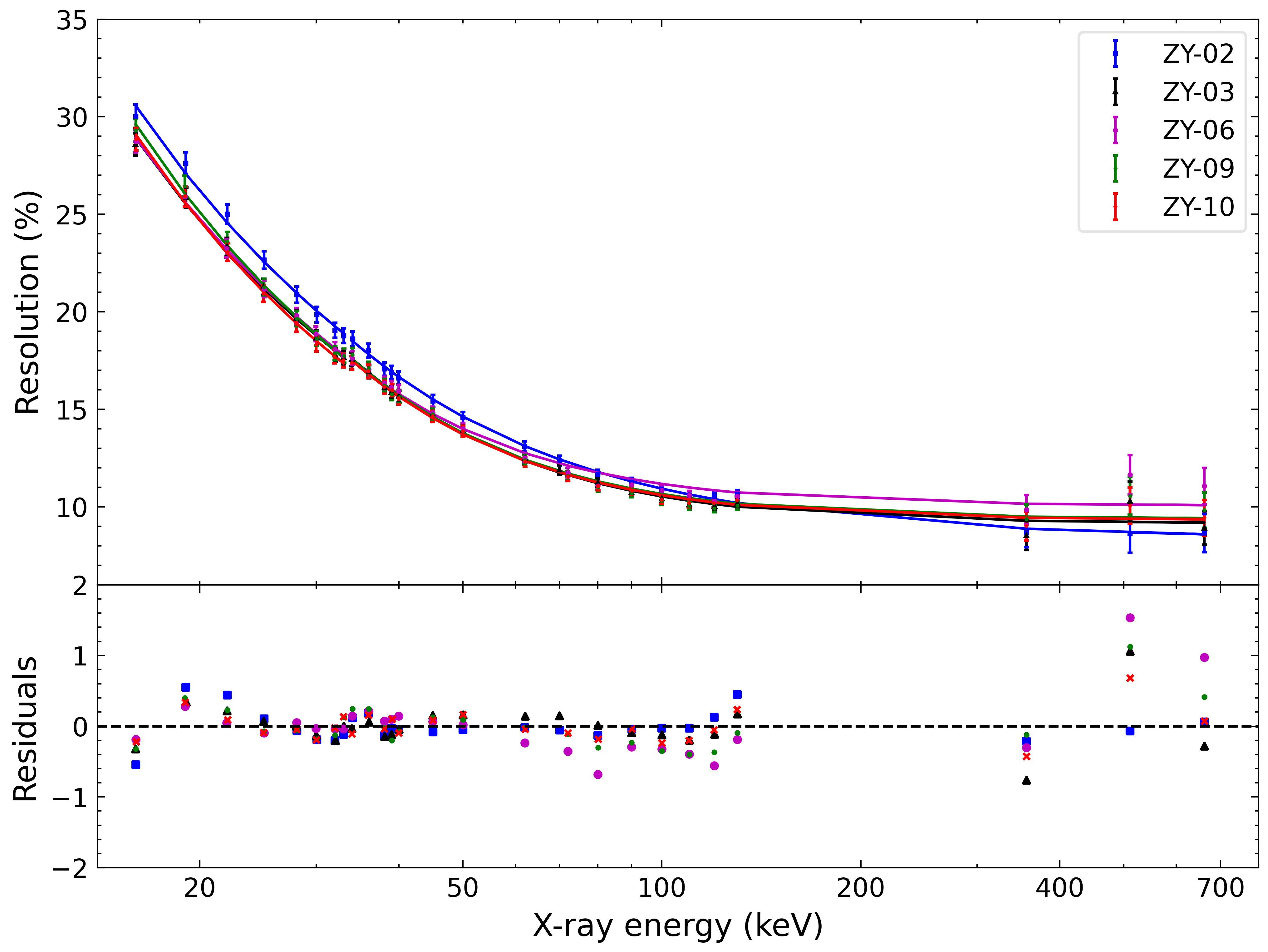}
    \caption{Energy–resolution relationships and fitted curves for five flight GTPs, with residuals obtained by subtracting the experimental values from the fitted ones.}
    \label{figure 14}
\end{figure}

The method described in Section~\ref{chap:Energy Spectrum and Fitting} was applied to fit the X-ray tube and radioactive source spectra from all the GTPs, yielding fitting results such as peak positions and standard deviations. The energy resolution of GTPs was calculated by dividing the FWHM of the full-energy peak by the peak position. The resolutions of the five-flight GTPs are shown in Fig.~\ref{figure 14}. The curves depicting the energy resolution as a function of energy, as fitted by Equation~\ref{eq:E-R Fit}, are also shown. In this equation, the constant term $a$ represents electronic noise, the second term $b$ accounts for the statistical fluctuations of scintillation photons and photoelectrons, and the third term $c$ reflects the intrinsic contribution of the scintillator, primarily stemming from luminescence nonproportionality \cite{ref35, ref36}.

\begin{equation}\label{eq:E-R Fit}
Resolution(E_\gamma) = \frac{2.355 \cdot \sigma(E_\gamma)}{Ch(E_\gamma)} = \frac{\sqrt{a^2 + b^2E_\gamma + c^2E_\gamma^2}}{E_\gamma}.
\end{equation}

The effect of the X-ray machine $\sigma_{X-mach}$ on GTP's energy resolution is indirectly determined using Equation~\ref{eq:X RS} using HPGe. Here, $\sigma_{beam}$ represents the broadening of the X-ray beam measured by the HPGe and $\sigma_{HPGe}$ is the energy resolution of the HPGe itself. GTP's intrinsic energy resolution $\delta_{int}$ must be corrected for the effect of the X-ray machine. In Equation~\ref{eq:intrinsic RS}, $\sigma_{GTP}$ denotes the standard deviation of the full energy peak of the net spectrum, which provides the intrinsic energy resolution after removing the effect of the X-ray machine.
Figure~\ref{figure 14} presents the residuals and errors for each energy point. Fitting errors were computed using error propagation formulas, and the error bars accounted for the effects of the statistics and GTP uniformity.

\begin{equation}\label{eq:X RS}
\sigma_{X-mach}^2 = \sigma_{beam}^2 - \sigma_{HPGe}^2
\end{equation}

\begin{equation}\label{eq:intrinsic RS}
\delta_{int} = 2.355 \cdot \sqrt{\sigma_{GTP}^2 - \sigma_{X-mach}^2}.
\end{equation}

\begin{figure}[H]
\centering
	\includegraphics[width=\hsize]{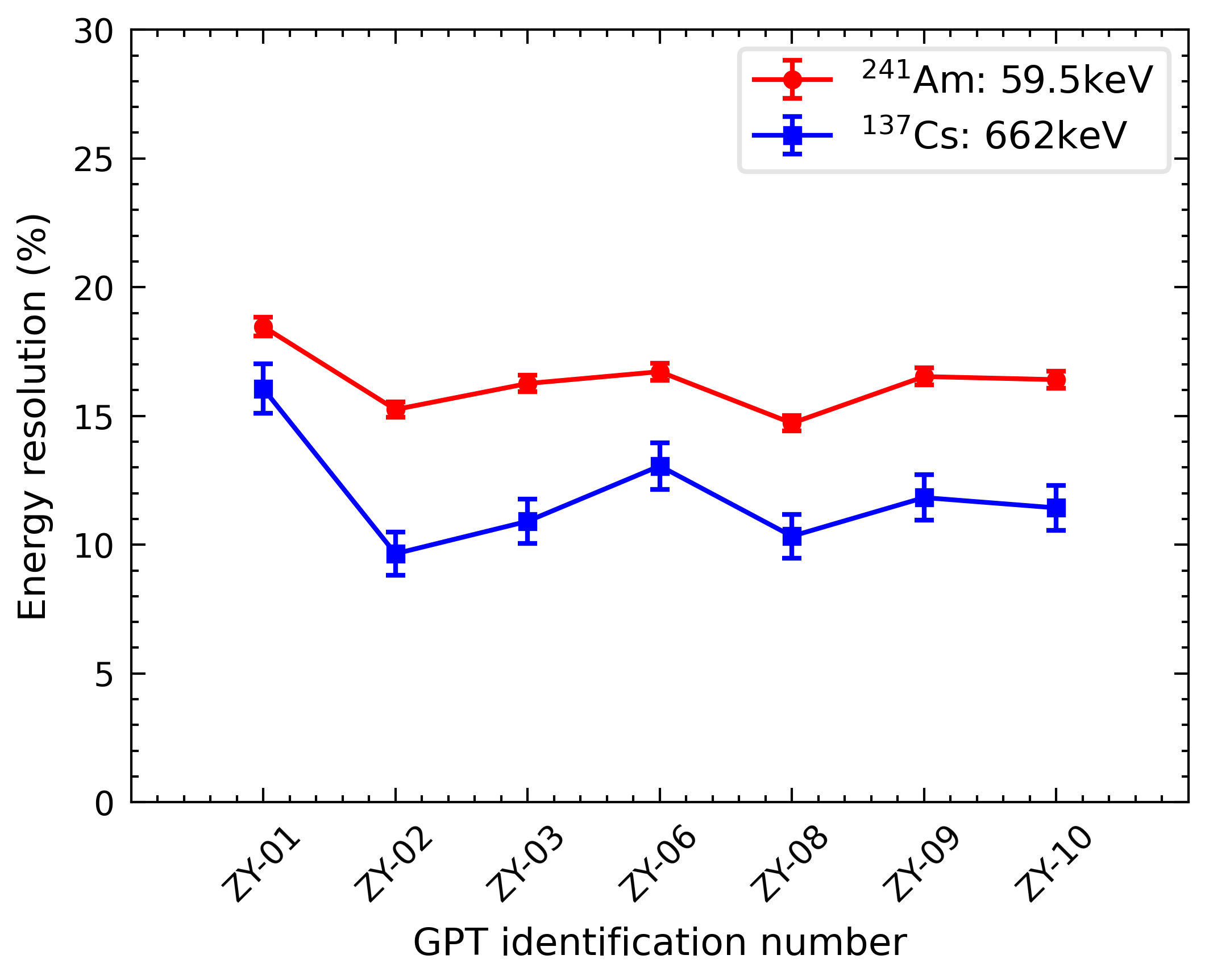}
    \caption{Energy resolution for 59.5 keV and 662 keV was measured for 5 flight GTPs and 2 backup GTPs, respectively.}
    \label{figure 15}
\end{figure}

The energy resolution is primarily determined by factors such as photoelectron statistics, nonuniform crystal luminescence, electronic noise, and the nonlinear response of the crystal. 
The NaI(Tl) crystals used in GTP have a large aspect ratio (width/thickness) and exhibit noticeable non-uniformity, resulting in poorer energy resolution compared to crystals with smaller aspect ratios.
High-energy X/$\gamma$-rays can either scatter in the GTP, deposit only a fraction of their energy, or penetrate the crystal directly without depositing energy. Both scenarios led to fewer detected high-energy events, resulting in deteriorated resolution. The high energy range shown in Fig.~\ref{figure 14} only presents three points tested using radioactive sources because of the inferior resolution of the NaI (Tl) crystal, making it impossible to distinguish certain energy peaks in the spectrum, with the highest usable full-energy peak at 662 keV. Figure~\ref{figure 15} illustrates the energy resolution of the main peak for each GTP using $^{241}$Am and $^{137}$Cs radioactive sources. For 59.5 keV $\gamma$-rays, all GTPs exhibit energy resolutions that meet the design target of less than 25\%.

\subsection{Detection Efficiency}\label{chap:Detection Efficiency}

\begin{figure}[H]
\centering
	\includegraphics[width=\hsize]{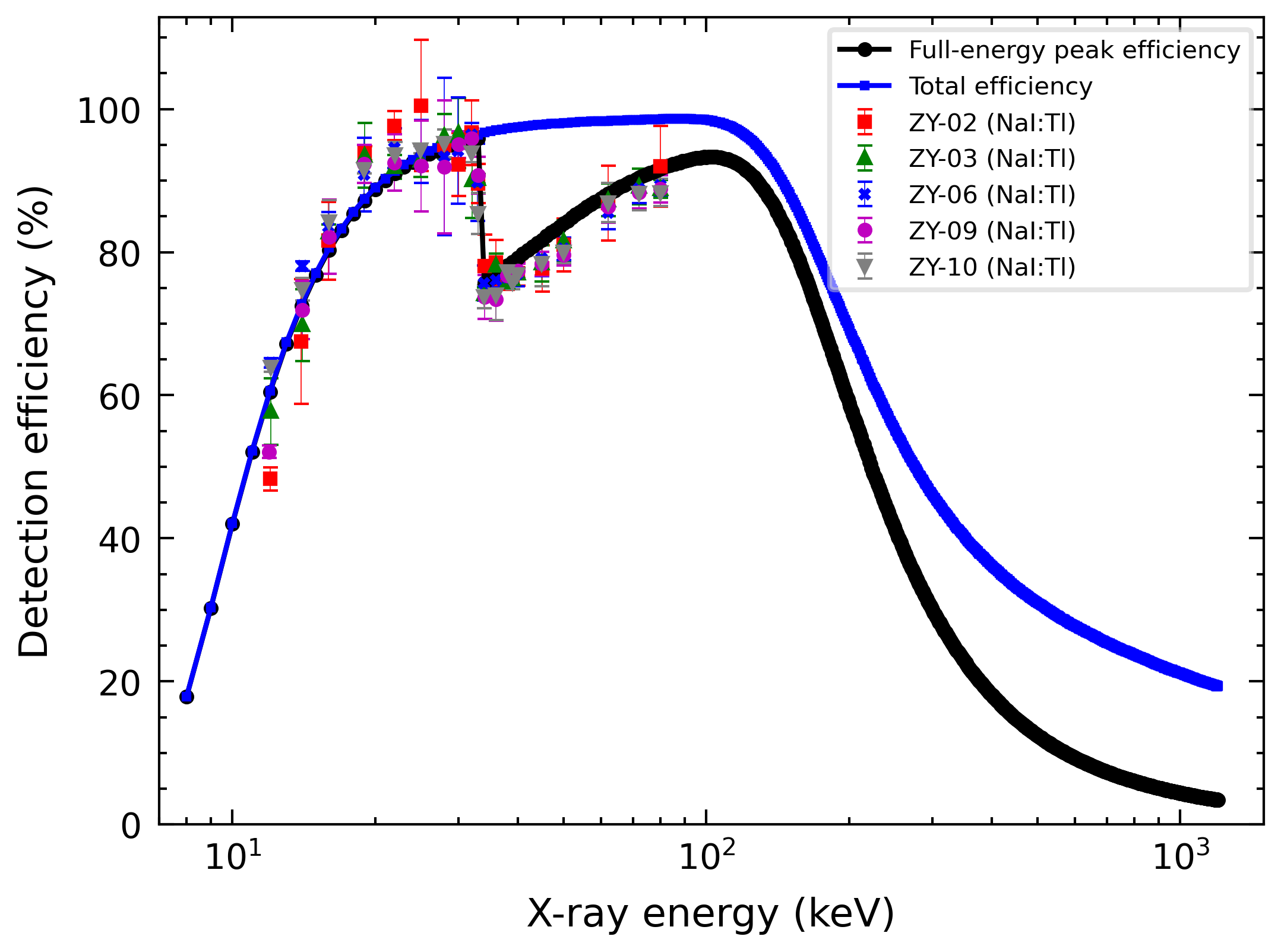}
    \caption{Comparison between experimentally measured detection efficiency and simulation results.}
    \label{figure 16}
\end{figure}

The detection efficiency of the GTP full-energy peak calibrated by the HXCF can be indirectly inferred from the efficiency of the HPGe detector \cite{ref37}. Equations~\ref{eq:beamstablity} and~\ref{eq:Efficiency} are used to calculate the full-energy peak detection efficiency, where $n_{GTP}$ and $n_{HPGe}$ denote the counts within 2.58$\sigma$ of the full-energy peaks. $I$ represents the intrinsic flux of the X-ray beams, $\varepsilon_{HPGe}$ indicates the full-energy peak efficiency of the HPGe detector, $t_{HPGe}$ is the testing duration for the HPGe detector (100 s), and $t_{GTP}$ represents the testing duration for the GTP (120 s). $\kappa_I$ denotes the beam stability obtained from the beam-monitoring system \cite{ref35}. Figure~\ref{figure 16} shows the intrinsic full-energy peak detection efficiencies of the five flight GTPs.

\begin{equation}\label{eq:beamstablity}
I(E_\gamma) = \frac{n_{HPGe}(E_\gamma)}{\varepsilon_{HPGe}(E_\gamma) \cdot t_{HPGe}}.
\end{equation}

\begin{equation}\label{eq:Efficiency}
\varepsilon_{GTP}(E_\gamma) = \frac{n_{GTP}(E_\gamma)}{I(E_\gamma) \cdot \kappa_I(E_\gamma) \cdot t_{GTP}}.
\end{equation}

To validate the accuracy of the quality model of the gamma-ray detector, Geant4 was employed to simulate the full-energy peak and total detection efficiencies. To maintain consistency with the ground calibration experiment, gamma photons were incident at the center position of the detector in a point-source form, and the ratio of the full-energy peak counts to the incident gamma photons was computed, resulting in a simulated full-energy peak detection efficiency. We also calculated the total detection efficiency by comparing the total spectral count with the number of incident gamma photons. 
As shown in Fig.~\ref{figure 16}, the simulation results for the low-energy range are consistent with the data from the ground calibration experiments. In the high-energy range, the GTP detection efficiency decreased significantly, particularly the detection efficiency of the full-energy peak. For 1 MeV gamma rays, the full-energy peak efficiency was 4.44\%, and the total detection efficiency was 21.24\%.

\subsection{Spatial Response}

\begin{figure}[H]
\centering
	\includegraphics[width=0.8\hsize]{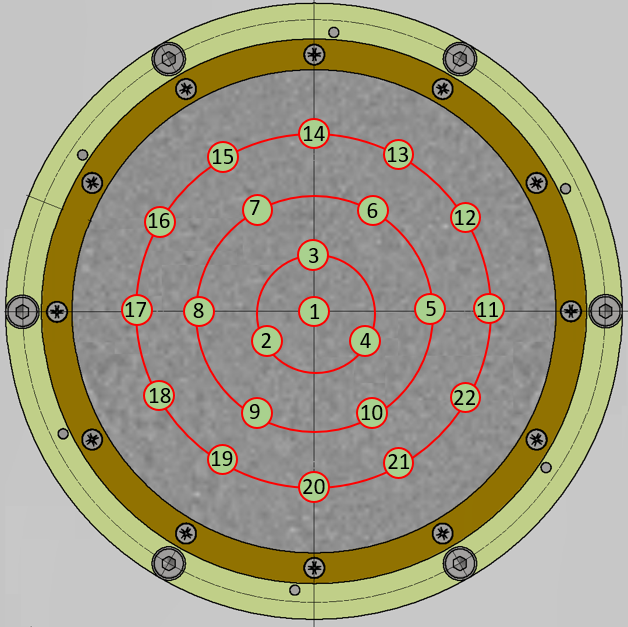}
    \centering
    \caption{The distribution and numbering of test points for assessing the positional non-uniformity of GTPs involve 22 designated locations.}
    \label{figure 17}
\end{figure}

\begin{figure}[H]
\centering
	\includegraphics[width=\hsize]{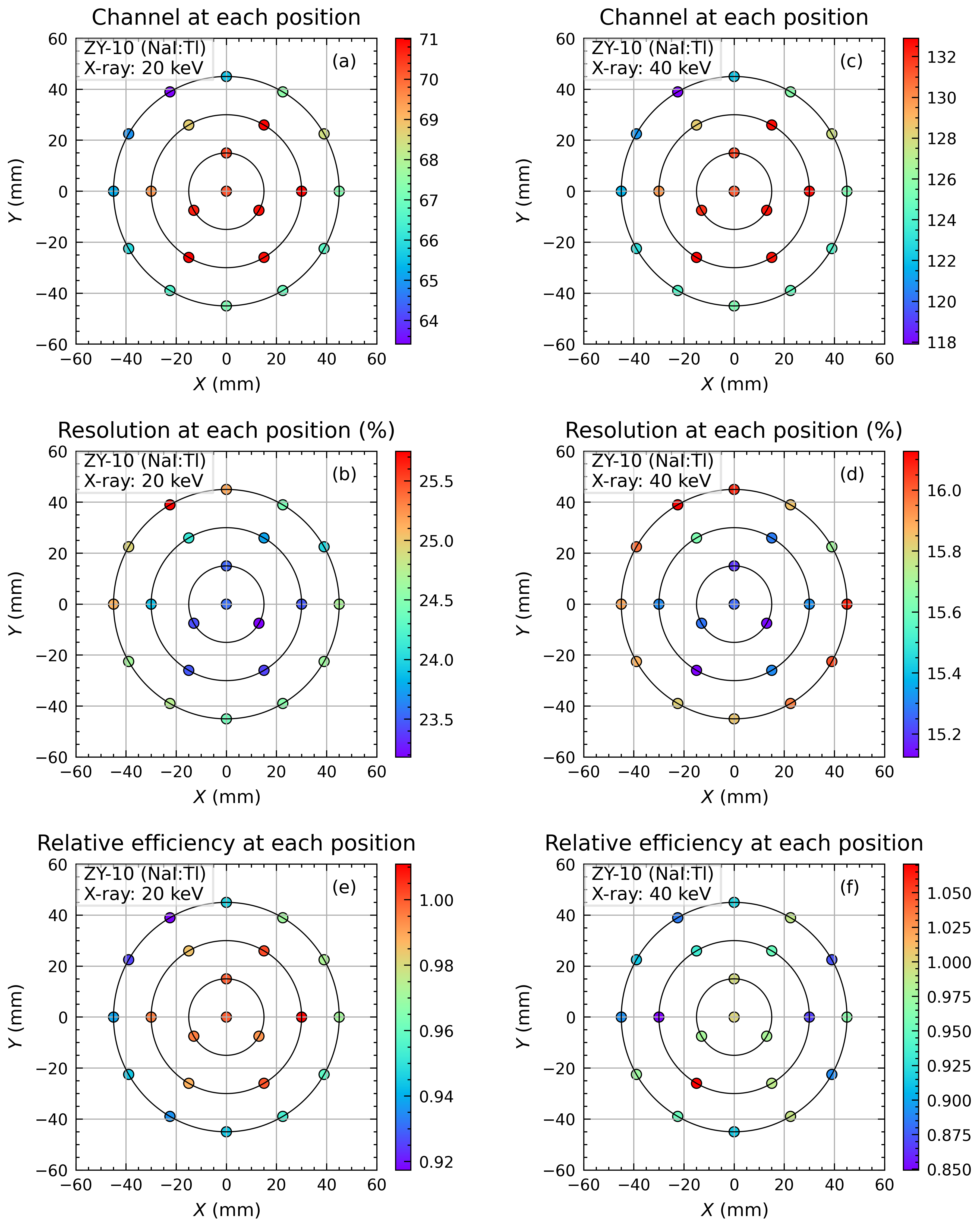}
    \caption{The positional non-uniformity results of ZY-10 GTP were tested using 20 keV (left) and 40 keV (right) X-rays, depicting variations in channel (top), energy resolution (middle), and relative detection efficiency (bottom) concerning different positions.}
    \label{figure 18}
\end{figure}

Owing to crystal luminescence nonuniformity and variations in light collection, GTPs show differing responses to X-rays at various positions. The position response testing of the GTPs was conducted using the HXCF by setting the parameter coordinates for each point and moving the displacement platform to align the X-ray beam with the positions to be tested. The incident X-ray positions are evenly distributed across the field of view, as shown in Fig.~\ref{figure 17}. Twenty-two numbered position points were evenly positioned along circles with radii of 15, 30, and 45 mm. X-ray energies of 20 keV and 40 keV were employed, with each point tested for 120 s, and the corresponding background data were collected for subtraction. Using the spectral fitting and efficiency calculation methods mentioned in Sections~\ref{chap:Energy Response} and~\ref{chap:Detection Efficiency}, the nonuniformity results were obtained, as illustrated in Fig.~\ref{figure 18}. The coordinates in Fig.~\ref{figure 18} were established with the GTP center (i.e., position point ‘1’ in Fig.~\ref{figure 17}) as the origin, where X and Y represent the azimuth and distance of the test points, respectively.

Position-response testing was performed for all flight and backup GTPs. ZY-08 and ZY-10 were tested with X-rays at 20 keV and 40 keV, whereas the other five GTPs were exposed to 59.5 keV $\gamma$-rays from a $^{241}$Am radioactive source. Taking ZY-10 as an example, Figure~\ref{figure 18} shows the peak channel, energy resolution, and relative detection efficiency at different positions. The relative standard deviations (RSDs) of these quantities were calculated using Equation~\ref{eq:RSD} to assess the nonuniformity of the GTPs. Additionally, nonuniformity can be quantified by calculating the maximum deviation (MD) between the GTP edges and the central point using Equation~\ref{eq:MD}. The nonuniformity test results for GTP ZY-10 are listed in Table~\ref{tab:non-uniformity test results}. We found that the non-uniformity was larger when tested with 40 keV X-rays compared with 20 keV. The RSD values remained within 6\%, indicating that ZY-10 exhibited good overall uniformity. The MD values were almost entirely within 10.21\%, but when tested with 40 keV X-rays, the MD for the relative detection efficiency reached 15.06\%. 
This suggested the presence of a pronounced edge effect in ZY-10, which was likely associated with the GTP response inhomogeneity. Furthermore, the X-ray machine flux has an instability of approximately 1\%, which can lead to deviations in detection efficiency.
Similar to ZY-10, the other six GTPs exhibited noticeable edge effects. The peak position, energy resolution, and detection efficiency at the edges were significantly lower than those at the central positions.

\begin{equation}\label{eq:RSD}
RSD = \frac{1}{\overline{x}}\sqrt{\frac{1}{n-1}\sum_{i=1}^n (x_i - \overline{x})^2}.
\end{equation}

\begin{equation}\label{eq:MD}
MD = max(\frac{x_i - x_{center}}{x_{center}}).
\end{equation}

\begin{table}[H]
	\centering
	\caption{Assess the positional non-uniformity of ZY-10 using the relative standard deviation (RSD) and the maximum edge-to-center deviation (MD).}
	\label{tab:non-uniformity test results}
 \begin{tabular*}{8.5cm} {@{\extracolsep{\fill} } >{\centering\arraybackslash}m{1cm}>{\centering\arraybackslash}m{1.2cm}>{\centering\arraybackslash}m{1.4cm}>{\centering\arraybackslash}m{1,4cm}>{\centering\arraybackslash}m{1.5cm}}
		\hline
		Energy & Quantities & Peak channel & Resolution & Relative efficiency\\
		\hline
		\multirow{2}{*}{20 keV} & RSD & 3.49\% & 2.89\% & 2.92\%\\
            & MD & 9.74\% & 9.34\% & 8.26\%\\
		\multirow{2}{*}{40 keV} & RSD & 3.63\% & 2.27\% & 5.81\%\\
            & MD & 10.21\% & 5.70\% & 15.06\%\\
		\hline
	\end{tabular*}
\end{table}

\subsection{Bias-voltage Response}\label{Bias-voltage Response}

At room temperature (20 $\pm$2 ℃), the bias-voltage response of two GTPs, ZY-06 and ZY-08, was investigated using $^{241}$Am and $^{137}$Cs radioactive sources. The SiPM bias voltage was incrementally adjusted in steps of 0.1 V within a voltage range of 25.2–27.2 V. Experimental data were corrected for dead time and background effects, and corresponding energy spectra were fitted to obtain trends in peak position and energy resolution with varying SiPM bias voltage.

\begin{figure}[H]
\centering
	\includegraphics[width=\columnwidth]{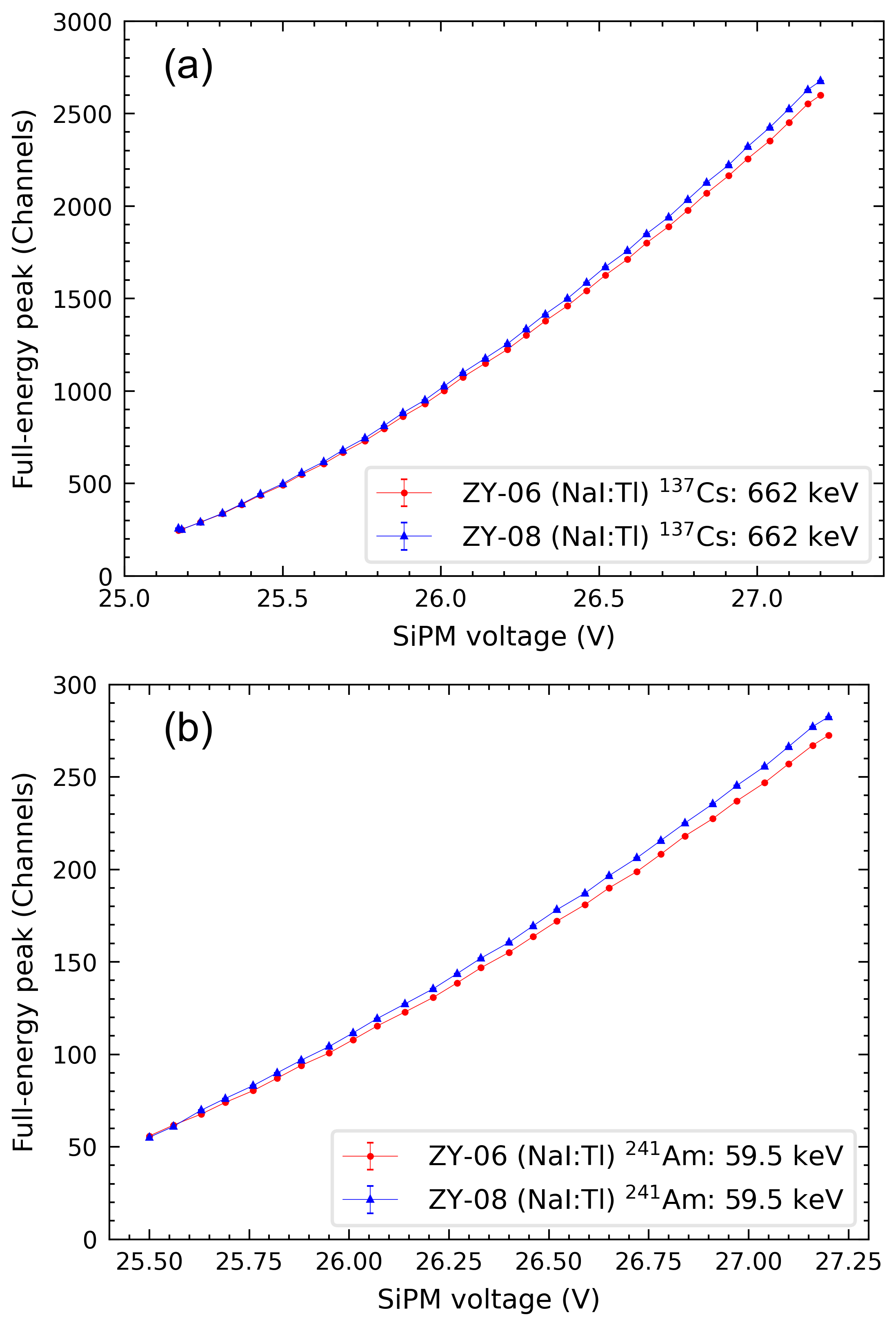}
    \caption{Testing ZY-06 and ZY-08 GTPs using the $^{137}$Cs (a) and $^{241}$Am (b) radioactive sources to observe the full-energy peak position changes concerning SiPM bias voltage.}
    \label{figure 19}
\end{figure}

\begin{figure}[H]
\centering
	\includegraphics[width=\columnwidth]{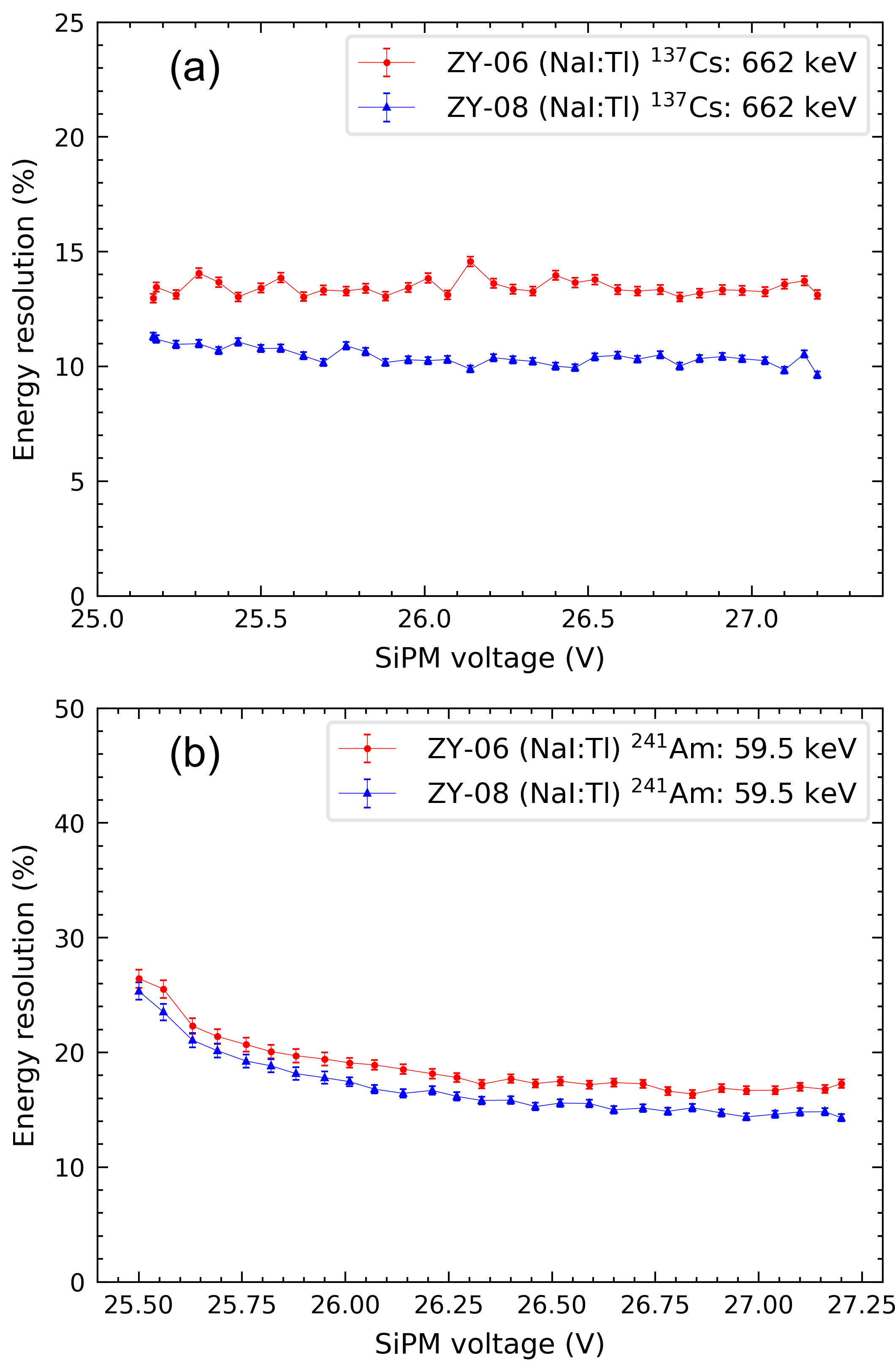}
    \caption{Testing ZY-06 and ZY-08 GTPs using the $^{137}$Cs (a) and $^{241}$Am (b) radioactive sources to observe the energy resolution changes of the full-energy peak concerning SiPM bias voltage.}
    \label{figure 20}
\end{figure}

The results for the full energy peak are presented in Fig.~\ref{figure 19}, which shows an exponential increase in the peak position with increasing voltage. The energy resolution results are shown in Fig.~\ref{figure 20}, revealing that, for 662 keV $\gamma$rays, the fluctuation in the energy resolution within the tested voltage range does not exceed 1.67\%. However, for 59.5 keV $\gamma$rays, voltages less than 25.5 V are constrained by the detection threshold, resulting in an incomplete full-energy peak; thus, the testing range was limited from 25.5 V to 27.2 V. After the GTP was placed in orbit, it exhibited a bias-voltage adjustment range. The energy resolution of the ground calibration remains nearly unchanged near a 26.5 V bias, affirming the suitability of selecting 26.5 V as the ground calibration bias voltage.

\subsection{Temperature Dependence}\label{Temperature Dependence}

After the launch of the GTM into orbit, a substantial temperature difference exists between the deep-space environment and the ground. To investigate the effects of temperature on the gain of the detector, it is essential to conduct ground temperature experiments on the GTP and establish a temperature-bias-voltage response matrix. The on-orbit temperature design values for the four standard GTPs shown in Fig.~\ref{fig:figure 1} range from –35 to –20 ℃, while for the –Z side GTP, the on-orbit temperature design value ranges from –35 to +20 ℃. In the ground experiments, GTP ZY-05 was placed in a High- \& Low-Temperature Chamber, covering a temperature range of –37 °C to 25 °C (comprising 14 temperature points), encompassing the on-orbit temperature design values of GTPs. Once the preset temperature was reached, the SiPM bias voltage was incrementally adjusted in the range of 25.3 V to 27.2 V ( 20 voltage points). Tests were conducted at these voltages using $^{241}$Am and $^{22}$Na radioactive sources.

\begin{figure}[H]
\centering
	\includegraphics[width=\columnwidth]{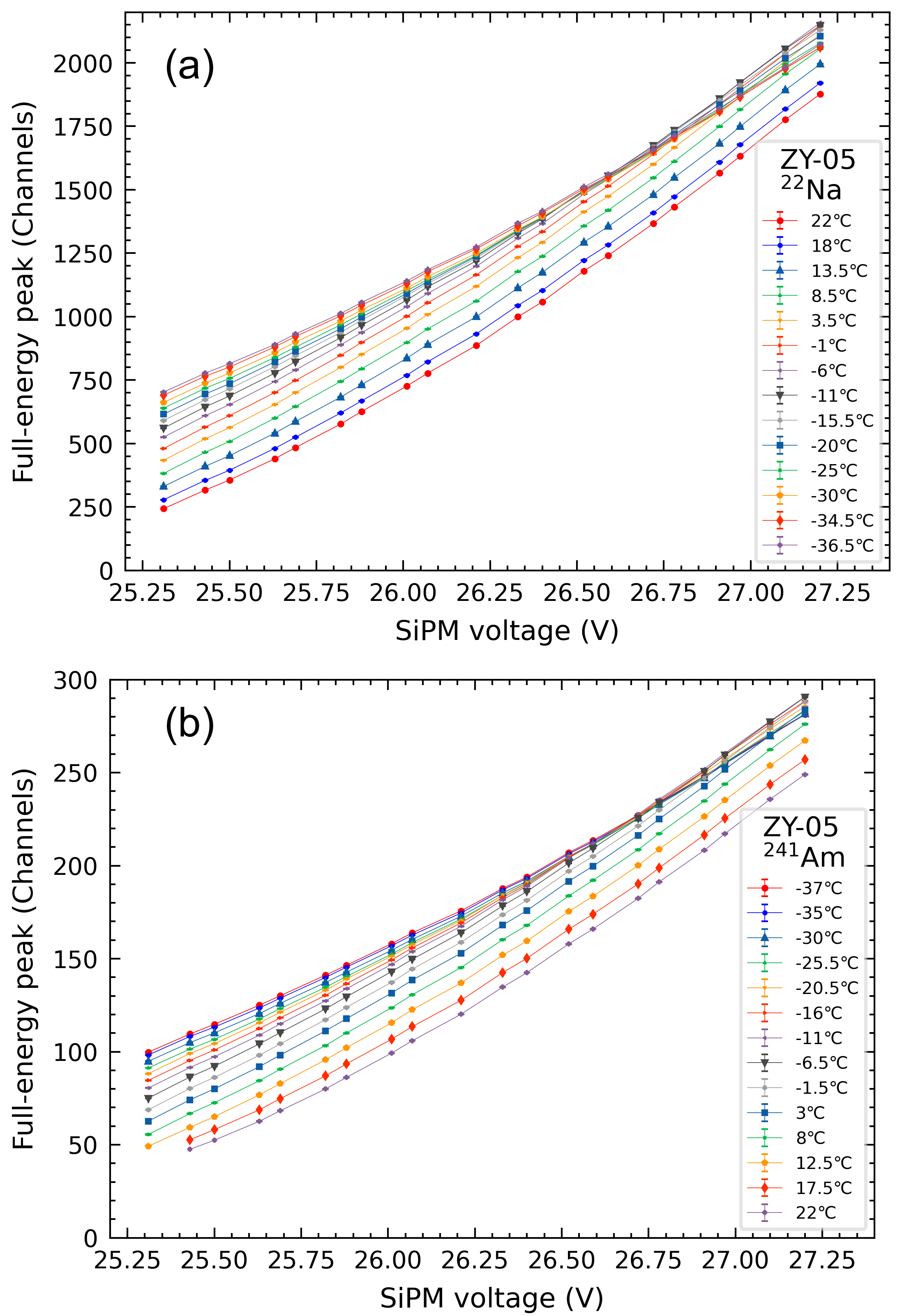}
    \caption{Testing ZY-05 GTP using the $^{22}$Na (a) and $^{241}$Am (b) radioactive sources at various temperatures to observe the variation in the full-energy peak position concerning SiPM bias voltage.}
    \label{figure 21}
\end{figure}

\begin{figure}[H]
\centering
	\includegraphics[width=\columnwidth]{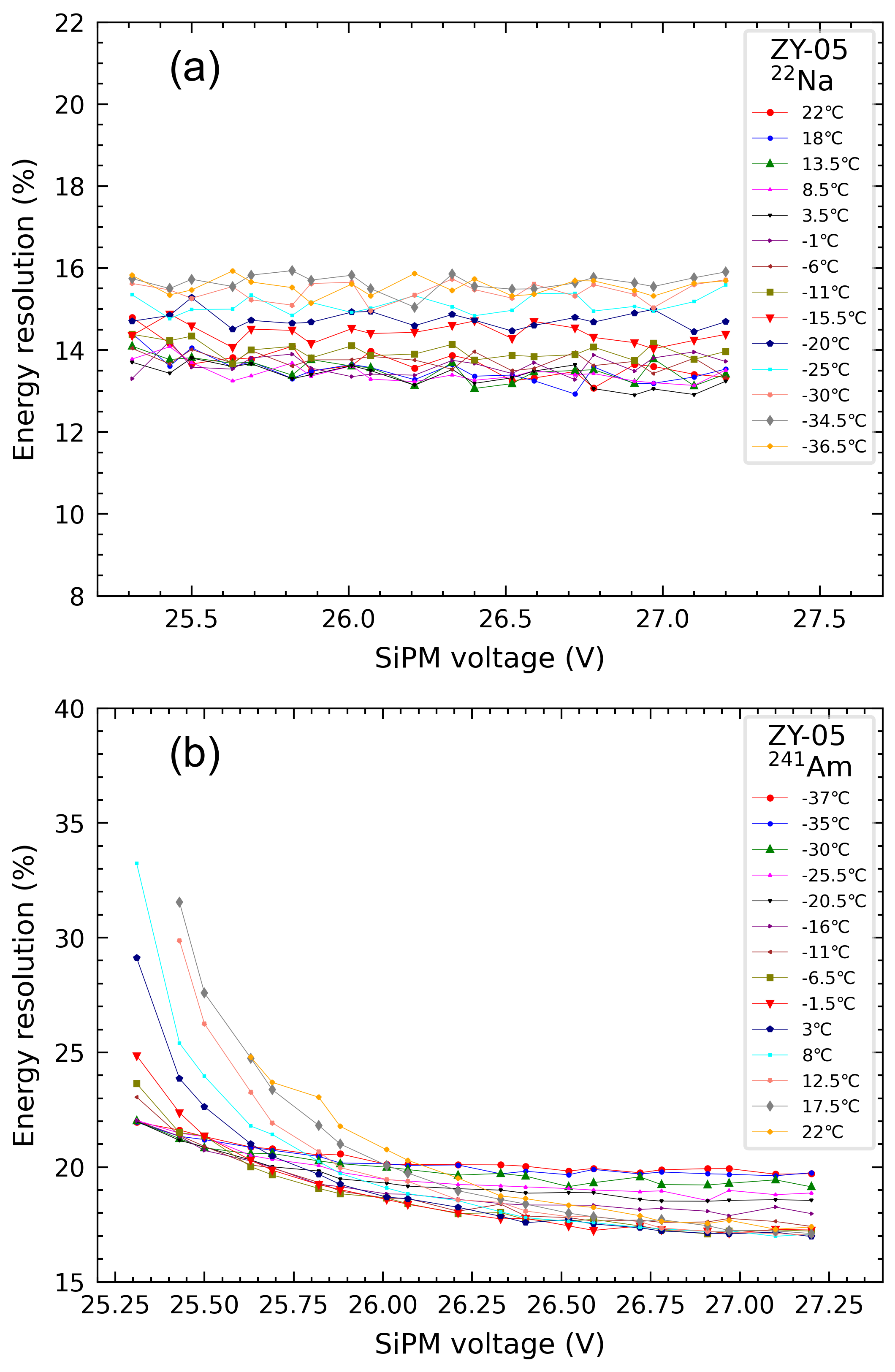}
    \caption{Testing ZY-05 GTP using the $^{22}$Na (a) and $^{241}$Am (b) radioactive sources at various temperatures to observe the variation in the energy resolution of full-energy peak concerning SiPM bias voltage.}
    \label{figure 22}
\end{figure}

The results for the peak position and energy resolution of GTP ZY-05 are shown in Fig.~\ref{figure 21} and Fig.~\ref{figure 22}. From these figures, it is evident that at lower temperatures, the changes in peak position and energy resolution with respect to voltage are more gradual, but there is a significant difference in gain compared to that at room temperature. Operating at lower voltages can address the radiation resistance issues of SiPMs in orbit, thereby extending the operational lifespan of the GTP while maintaining its performance. Based on the current test results, it is recommended to set the SiPM voltage to around 26 V when the environmental temperature is –30 °C.

Based on the results of the temperature experiments, we observed an evident temperature dependence in the SiPM-based gamma-ray detector. During the operation of the GTM payload, the temperature varies with changes in the satellite orbit. To ensure real-time gain stability and consistency of the GTPs, we employed a gain correction method that was successfully applied to GECAM-A/B and GECAM-C \cite{ref23, ref38}. This method allows real-time updating of the SiPM bias voltage based on detector temperature.

\begin{figure}[H]
\centering
	\includegraphics[width=\columnwidth]{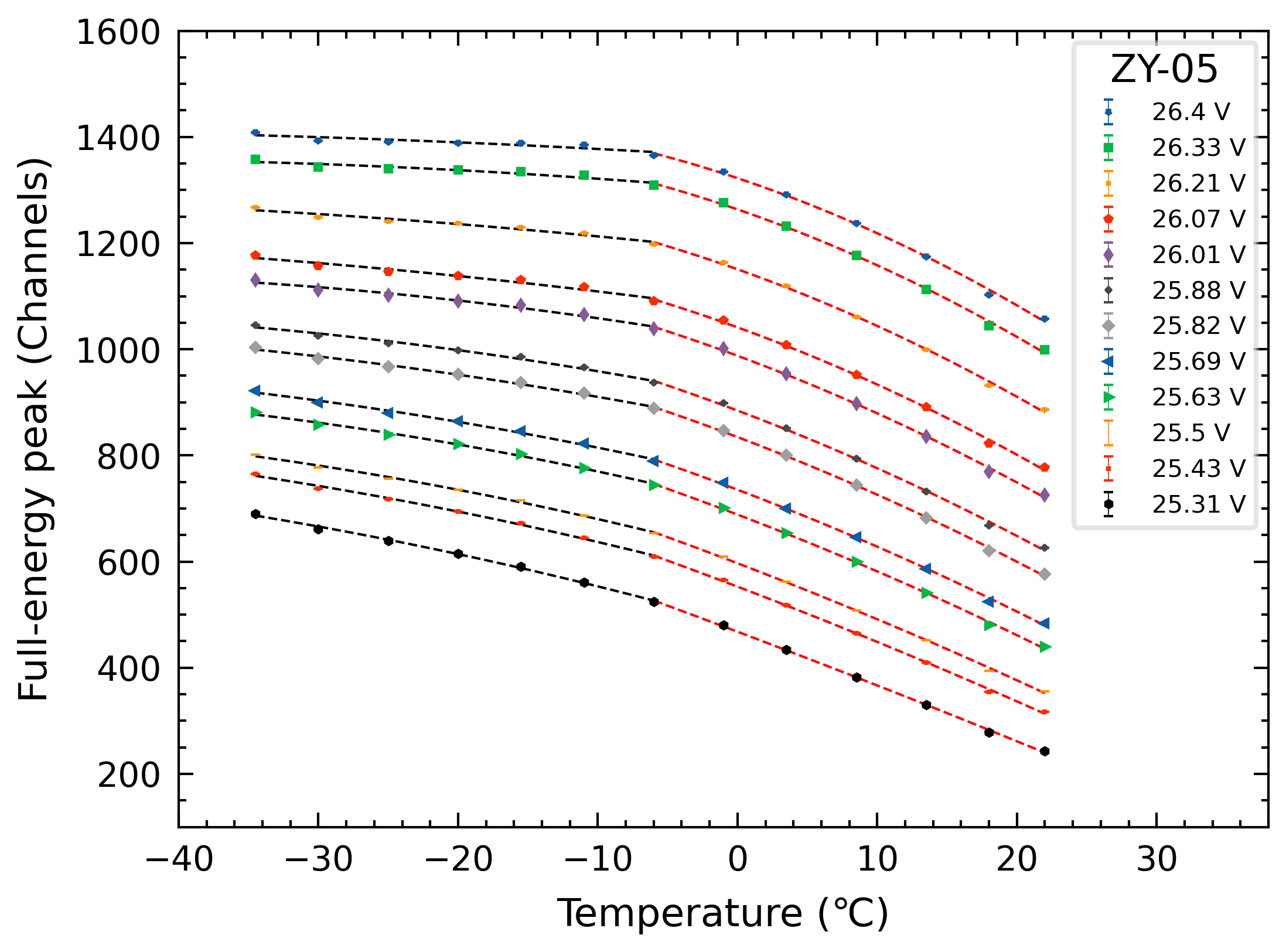}
    \caption{Testing ZY-05 GTP using a $^{22}$Na radioactive source at different SiPM bias voltages to observe the peak position changes of the 511 keV full-energy peak, illustrating the temperature dependency of the gamma-ray detector.}
    \label{figure 23}
\end{figure}

To establish a flexible temperature–voltage lookup table (LUT), it is necessary to determine the temperature dependence coefficient (P$_{td}$) of the gamma-ray detector. The gain drift of the GTP was measured using 511 keV $\gamma$-rays from a $^{22}$Na radioactive source, covering a temperature range of –35–22 °C and a bias voltage range of 25.31–26.4 V. NaI(Tl) crystal exhibits negative temperature dependence \cite{ref39} but its temperature effect is non-linear, with light yield decreasing at both low and high temperatures. The SiPM has a constant positive temperature dependence \cite{ref40}, and the temperature dependence coefficient of GTP is a combined effect of NaI(Tl) and the SiPM. Figure~\ref{figure 23} shows the trend of the 511 keV energy peak position (Chn) with temperature (T). To facilitate subsequent calculation of the temperature coefficient, we employed Equation~\ref{eq:temperature fit} to fit the data points separately within the temperature ranges of –35 °C to –6 °C and –6 °C to 22 °C.

\begin{equation}\label{eq:temperature fit}
Chn = a + b \cdot T + c \cdot T^2.
\end{equation}

\begin{equation}\label{eq:temperature coefficient}
P_{td}(V_b) = \frac{V_b - V_0}{T_b(V_b,Chn_b) - T_0(V_0,Chn_b)}.
\end{equation}

\begin{figure}[H]
\centering
	\includegraphics[width=\columnwidth]{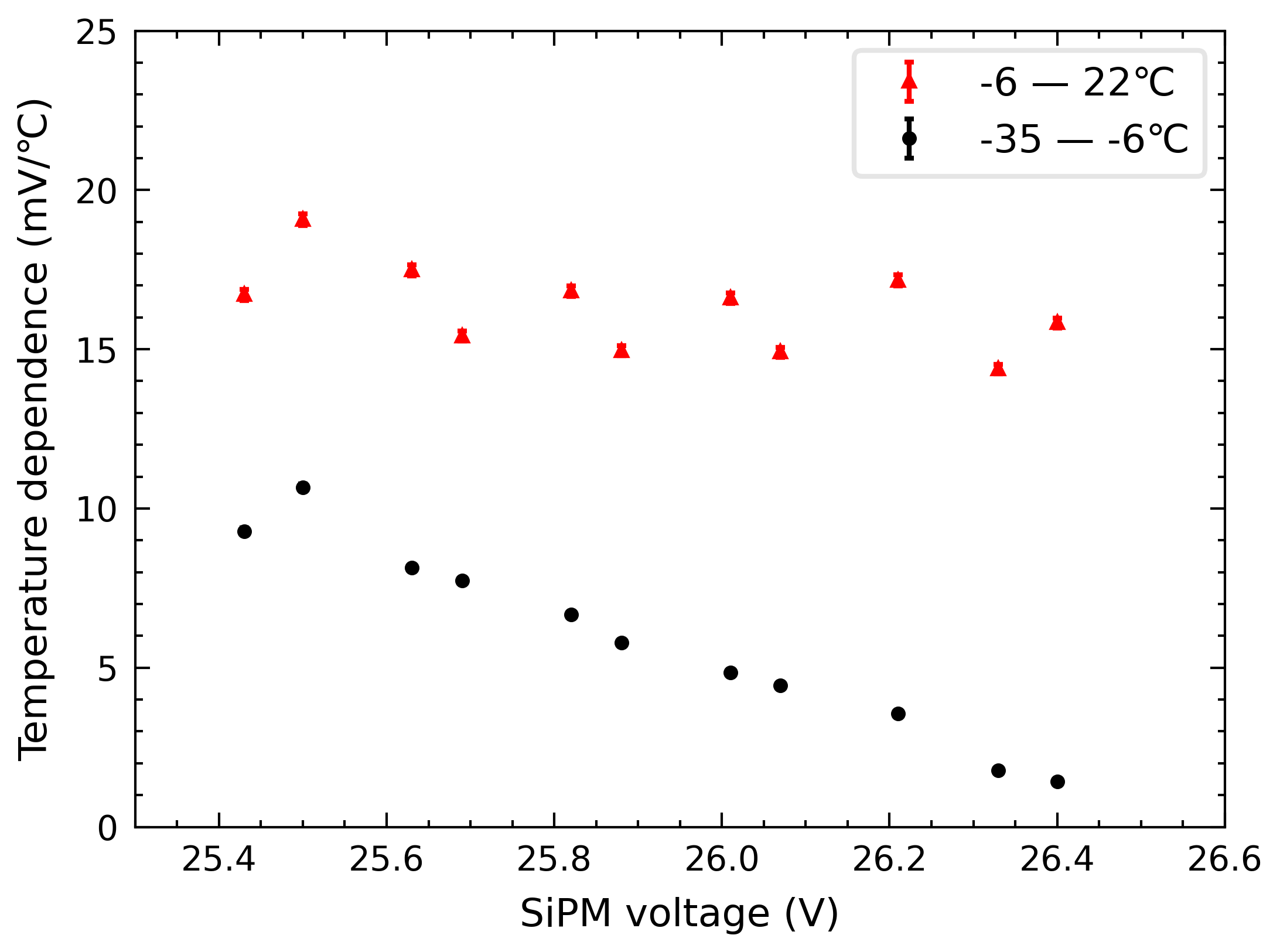}
    \caption{Temperature coefficients at various SiPM bias voltages.}
    \label{figure 24}
\end{figure}

During the actual in-orbit operations, the data acquisition system scans the temperature monitor (T) on the GTP every second. If the temperature changes exceed 0.5 °C, the SiPM bias voltage (V$_b$) is updated based on the temperature–voltage LUT. The definition of the temperature dependence of the GTP is similar to that of the GRD for GECAM-A/B and GECAM-C \cite{ref38}. V$_0$ represents the selected reference SiPM bias voltage, which is the lowest and closest bias voltage to V$_b$ in the experimental data. Chn$_b$ denotes the channel of the 511 keV peak position of the GTP at T$_b$ and V$_b$. In the two data segments shown in Fig.~\ref{figure 23}, T$_b$ is chosen as –6 °C and 22 °C. T$_0$(V$_0$, Chn$_b$) represents the temperature at V$_0$ with the same Chn$_b$ channel as determined from the curve in Figure~\ref{figure 23}. The temperature-dependence coefficients for each voltage, calculated using Equation~\ref{eq:temperature coefficient}, are shown in Fig.~\ref{figure 24}, indicating an overall positive temperature dependence of the GTPs.

\begin{figure}[H]
\centering
	\includegraphics[width=\columnwidth]{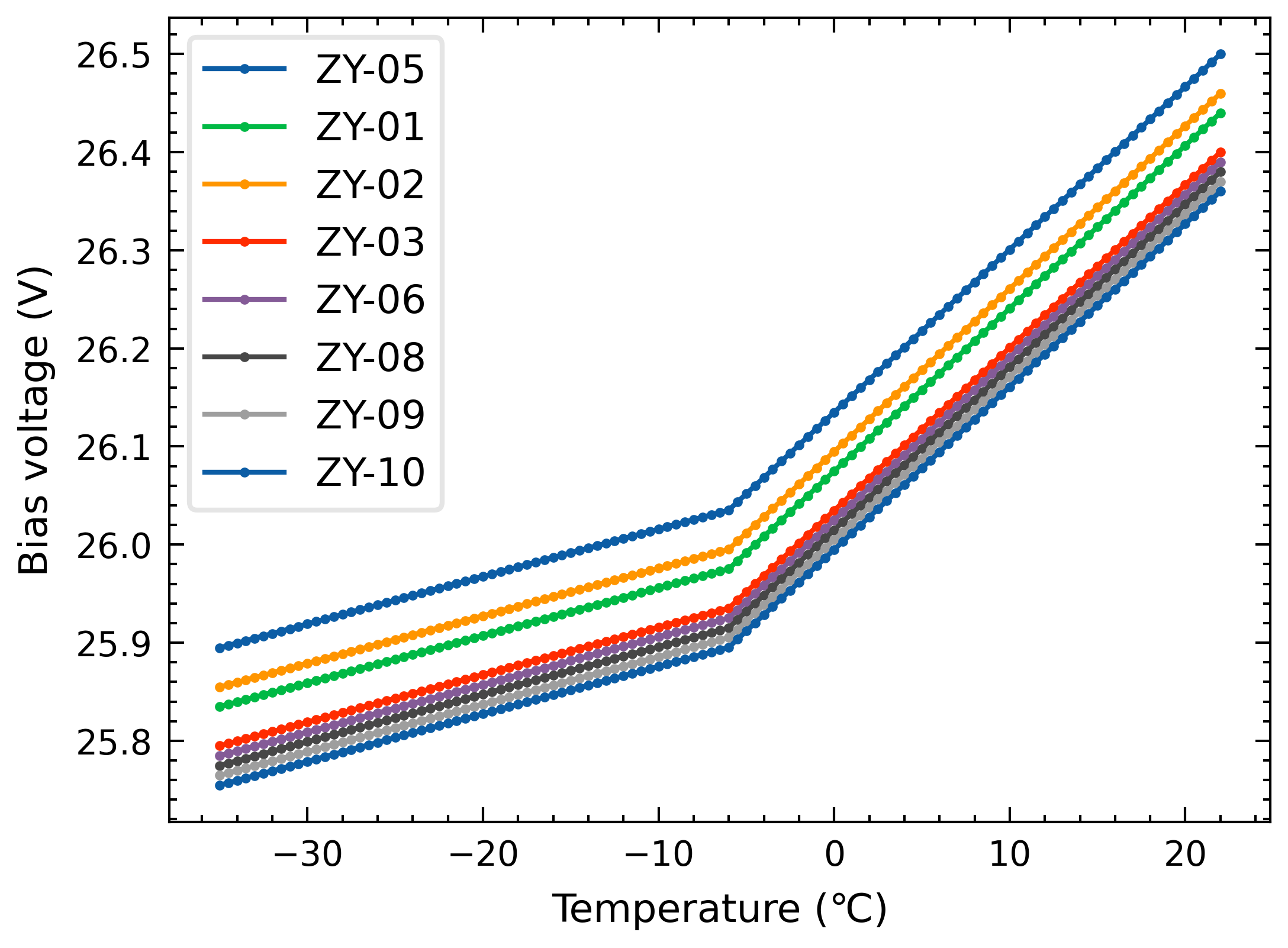}
    \caption{Temperature–voltage look-up table (LUT).}
    \label{figure 25}
\end{figure}

The operational voltage of the GTPs in orbit was set to approximately 26 V. According to Fig.~\ref{figure 24}, GTP exhibits a temperature coefficient of 4.84 mV/℃ within –35 to –6 ℃ and 16.60 mV/℃ within –6 to 22 ℃. Using the 511 keV ($^{22}$Na radioactive source) peak position of ZY-05 at room temperature as a reference, the temperature coefficient provides an appropriate SiPM bias voltage for each GTP at the same temperature (Fig.~\ref{figure 25}). This temperature compensation design ensures the stability and consistency of the GTP gains. 

Temperature-dependent correction is an important part of in-orbit calibration tests. We first applied ground-estimated LUTs in orbit and adjusted the temperature-dependence correction according to the in-orbit data. The embedded $^{241}$Am radioactive source in the GTP provides an energy line of 59.5 keV for in-orbit calibration. With the satellite experiencing temperature fluctuations post-launch, we will derive the relationship between the peak positions of the 59.5 keV energy line for each GTP over the satellite's in-orbit lifetime. Monthly monitoring of the LUTs will be conducted throughout the instrument's operational life, with adjustments made as necessary.

\section{Summary}

The GTM is a new all-sky gamma-ray monitor that will be launched into the DRO orbit in 2024. GTPs utilize large-area NaI(Tl) crystals as their detection-sensitive material coupled with SiPM arrays to create a novel design for a dual-channel real-time signal coincidence readout to suppress SiPM noise. In this study, we first investigated the impact of the coincidence time window and then conducted a comprehensive ground calibration of the GTP detector of DRO/GTM using the HXCF and radioactive sources, including energy–channel relationship, energy–resolution relationship, detection efficiency, spatial non-uniformity, bias-voltage response, and temperature experiments. The results indicate that all GTPs involved in ground calibration meet the expected specifications and the GTPs cover an energy range of 9 keV–1.1 MeV and perform well in the low-energy range. Ground calibration also validated the mass model of the detector and Monte Carlo simulation results of the detector response—the fundamental steps in the development and operation of GTM.

In addition, considering the significant differences between in-orbit environmental temperatures (approximately -30 °C) and ground calibration temperature (approximately 20 °C), the ground calibration results of energy response cannot be directly applied to deep-space DRO orbit and should take the temperature effect into account. Therefore, the temperature dependence was comprehensively tested in this study. After the launch of the GTM, a combination of ground and in-flight calibrations is required to establish a final calibration database.

\Acknowledgements{This work is supported by the Strategic Priority Research Program of the Chinese Academy of Sciences (Grant No. XDA30050100 and XDA30030000) and the National Natural Science Foundation of China (Grant Nos. 12173038, 11775251, 12273042, and 12075258). The GECAM (Huairou-1) mission was funded by the Strategic Priority Research Program on Space Science (XDA15360000) of the Chinese Academy of Sciences (CAS). We thank the staff of the Shandong Institute of Aerospace Electronic Technology, National Institute of Metrology, for significantly helping in the development and ground calibration tests of the GTM. We also thank the DRO team.}

\InterestConflict{The authors declare that they have no conflict of interest.}

\end{multicols}
\end{document}